\documentclass{ws-p9-75x6-50}

\usepackage{overcite}

\def\beq{\begin{equation}} \def\eeq{\end{equation}}
\def\art#1{#1,}

\input prepictex \input pictex \input postpictex
\def\mathput#1{\relax \ifmmode \displaystyle #1\else $\displaystyle #1$\fi}

\newcommand{\pubheading}[1]
{\vspace*{-7.5cm}\vspace*{-12pt}
\vbox to 0pt{
  {\flushleft
	{\footnotesize Proceedings of the Ninth ICRA Network Workshop 
   \hfill
}\\[-2pt]
  	{\footnotesize 
on Fermi and Astrophysics (2001)}\\[-2pt]
  	{\footnotesize
 Eds.\ R. Ruffini and C. Sigismondi #1 }\\[-2pt]
	{\footnotesize 
\copyright 2003 World Scientific}\\
	} \vss}
\vspace*{7.5cm}\vspace*{12pt}\nointerlineskip} 

\def\rightnote{INW9 Proceedings (2003): {\it Jantzen}}%

\begin{document}

\title{Circular Holonomy, Clock Effects and Gravitoelectromagnetism: Still Going Around in Circles After All These Years $\ldots$}

\author{DONATO BINI}

\address{Istituto per le Applicazioni del Calcolo ``M. Picone", C.N.R.,     
I--00161 Rome, Italy\\
and I.C.R.A., Universit\`a di Roma, I--00185 Roma, Italy\\
e-mail: binid@icra.it}
\author{ROBERT T. JANTZEN}

\address{Department of Mathematical Sciences, Villanova University, 
  Villanova, PA 19085, USA\\
and I.C.R.A., Universit\`a di Roma, I--00185 Roma, Italy \\
e-mail: robert.jantzen@villanova.edu}

\maketitle

\pubheading{}  

\abstracts{
The historical origins of Fermi-Walker transport 
and Fermi coordinates 
and the construction of Fermi-Walker transported frames in black hole spacetimes are reviewed. For geodesics this transport reduces to parallel transport and these frames can be explicitly constructed using Killing-Yano tensors as shown by Marck. For accelerated or geodesic circular orbits in such spacetimes, both parallel and Fermi-Walker transported frames can be given, and allow one to study circular holonomy and related clock and spin transport effects. In particular the total angle of rotation that a spin vector undergoes around a closed loop can be expressed in a factored form, where each factor is due to a different relativistic effect, in contrast with the usual sum of terms decomposition. Finally the Thomas precession frequency is shown to be a special case of the simple relationship between the parallel transport and Fermi-Walker transport frequencies for stationary circular orbits.}

\section{Introduction}

The geometry of circular orbits in general relativity is so rich that after all these years in which various aspects of it have been studied in many approaches, remarkably there is still something interesting left to say on the matter. Here we describe parallel and Fermi-Walker transport along these curves in black-hole spacetimes after reviewing the historical origins of Fermi-Walker transport\cite{fermi3,walker} and the other interesting class of curves for which the transport equations can be explicitly solved: general geodesics in stationary axisymmetric spacetimes admitting a Killing-Yano tensor,\cite{marck1,marck2,marck3}
a case which includes black hole spacetimes.
For circular orbits in black hole spacetimes, the total parallel transport or Fermi-Walker angle per orbital revolution is represented in factored form,\cite{bjm2002} revealing each of the spacetime geometric contributions to the final result, in contrast with the usual sum of terms decomposition associated with the gravitoelectric (GE), gravitomagnetic (GM) and space curvature effects.\cite{mfg,idcf2}

Born in 1901, Fermi was a boy genius who grew up in Rome, arriving at the University of Pisa already having learned physics and advanced mathematics on his own, and having done many physics experiments with his friend Enrico Persico during high school (including measuring the precise acceleration of gravity in Rome) who together became the first two professors of theoretical physics in Italy in 1927.\cite{gillispie}
Fermi's first three papers (on electromagnetism and relativity) were  published in 1921\cite{fermi1,fermi2}  and 1922\cite{fermi3} while he still a university student at the University of Pisa caught up in the 
excitement of the newly born theory of general relativity (1916) and early evolution of quantum mechanics (he was the authority on the latter in Pisa as a student!), after which he returned to Rome briefly. Tullio Levi-Civita had just introduced the notion of parallel transport in 1917\cite{levicivitapt} (when Fermi finished high school) and had come to Rome from Padua in 1918 as a full professor of mathematics after having made fundamental contributions to the `absolute differential calculus'\cite{levibook} largely developed by his mentor Gregorio Ricci-Curbastro (hence the title `Ricci-Calculus' of Schouten's encyclopedic book in 1954\cite{schouten}) but improved and applied to physics by Levi-Civita and passed on to Einstein by Marcel Grossmann. More details on these historical figures can be found in the many volumes of Gillispie's standard reference.\cite{gillispie}  

Starting from Levi-Civita's form of the metric for a uniform gravitational field published in a series of articles of 1917--1918\cite{levicivita}
$$
 ds^2= -a^2 dt^2+ \delta_{ij} dx^i dx^j\ ,
$$
Fermi studied the effects of acceleration on Maxwell's equations in his second paper\cite{fermi2} (essentially equivalent to a `variable speed of light' $a$). He was then led to understand the way in which one could apply these results to a small enough spacetime region around the world line of a test observer in a first order approximation in the spatial distance from the world line, resulting in his third paper on what later came to be known as `Fermi transport.'  
Strangely making no mention of Fermi's earlier work and citing only Eisenhart\cite{eisenhart} (who does not discuss Fermi's work but does list his article in the bibliography) as a reference for Riemann normal coordinates (Eisenhart, appendix 3, see also section 11.7 of \citen{mtw}), Walker\cite{walker} repeated Fermi's calculations for transporting vectors taking a different approach, making explicit the implicit spacetime Fermi transport equation contained in Fermi's paper, and extended the analysis of the affect of curvature on the arclength of nearby curves to second order in the approximation, calculated only to first order by Fermi. Both examined the question on an $n$-dimensional manifold as an exercise in differential geometry. Eisenhart himself had extended Fermi's discussion a few years after it had appeared to a general symmetric (not necessarily metric) connection.\cite{eisenhart27}

\section{Geometric Overview}

In a spacetime with coordinates $x^\alpha$ and metric $g_{\alpha\beta}$ (signature $-$+++), consider an arbitrary timelike world line or spacelike curve  $x^\alpha(s)$ parametrized 
by an arclength paramter $s$ (respectively proper time and proper distance). The unit tangent is $u^\alpha = d x^\alpha /ds$ 
(4-velocity in the timelike case $\epsilon =-1$), where 
$ u\cdot u = u_\alpha u^\alpha = \epsilon$. The second derivative 
$D^2 x^\alpha/ds =DU^\alpha/ds = a^\alpha$ 
(acceleration when $\epsilon =-1$) then satisfies 
$a \cdot u=0$ (as follows from covariant differentiating $u\cdot u =\epsilon$ along the curve).
If $v$ is any vector defined along the curve which is orthogonal to $u$, i.e. $u\cdot v =0$, its equation of Fermi transport along the curve together with the transport equation for $u$ itself are
\beq
\frac{D u^\alpha}{ds} = a^\alpha\ ,\qquad
\frac{D v^\alpha}{ds} = -\epsilon u^\alpha (a_\beta v^\beta)\ , 
\eeq 
which were given by Fermi, so the second equation is often referred to as describing Fermi transport, generalizable to any tensors which are orthogonal to $u$. Vectors transported in this way along the curve remain orthogonal to $u$.

However, implicit in Fermi's work is the way to transport any vector $X$ along the curve, namely when decomposed orthogonally with respect to $u$, the part along $u$ is held constant while the orthogonal piece is evolved with this law. Letting
\beq
 X = X^{(||)} u + X^{(\bot)}\ , \quad
X^{(||)} = \epsilon X \cdot u\ , \quad
 X^{(\bot)} \cdot u = 0\ ,
\eeq
then if $X^{(||)}$ is held constant along the curve and $X^{(\bot)}$ is transported by the Fermi transport law, then since 
$X\cdot a = X^{(\bot)} \cdot a$
\begin{eqnarray}
\frac{D X^\alpha}{ds} 
&=& X^{(||)}  \frac{D u^\alpha}{ds} + u^\alpha \frac{D X^{(||)}}{ds}
 + \frac{D X^{(\bot)\alpha}}{ds}\ ,\nonumber\\
&=& \epsilon (a^\alpha u_\beta     - u^\alpha a_\beta) X^\beta\ ,
\end{eqnarray}
which is the general transport law given by Walker, subsequently referred to as Fermi-Walker transport, corresponding to vanishing Fermi-Walker derivative along the curve 
\beq\label{eq:fwtransport}
\frac{D_{\rm(fw)} X^\alpha}{ds}
=\frac{D X^\alpha}{ds}
-\epsilon (a^\alpha u_\beta     - u^\alpha a_\beta) X^\beta =0\ ,
\eeq
easily extended to any tensor defined along the curve. Relative to parallel transport along the curve, the additional terms
$-[\epsilon a\wedge u]^\alpha{}_\beta X^\beta$  
generate a rotation/pseudorotation in the plane of $u$ and $a$ by just the amount necessary to keep $u$ tangent to the curve, and for geodesics where $a=0$, this reduces to parallel transport. The metric itself has vanishing Fermi-Walker derivative along any curve and so in addition to preserving the orthogonal decomposition of the tangent space parallel and perpendicular to $u$ along the curve, all inner products are invariant under this transport.

Fermi derived his transport law by considering first an $n$-dimensional Riemannian manifold and then specialized his result to $n=4$ and scaled three coordinates by $i$ to change the signature to +$-$$-$$-$. His calculation can be retraced in modern language for the two cases $\epsilon =\pm1$ simultaneously. Although Fermi described the situation in words without any figures, a picture is worth a thousand words. Fig.~1 illustrates the argument for the simpler case of $a$ lying in the plane of $u$ and an orthogonal unit vector $v$, showing the ``infinitesimal" vectors associated with two infinitesimally separated points on the curve.

\begin{figure}
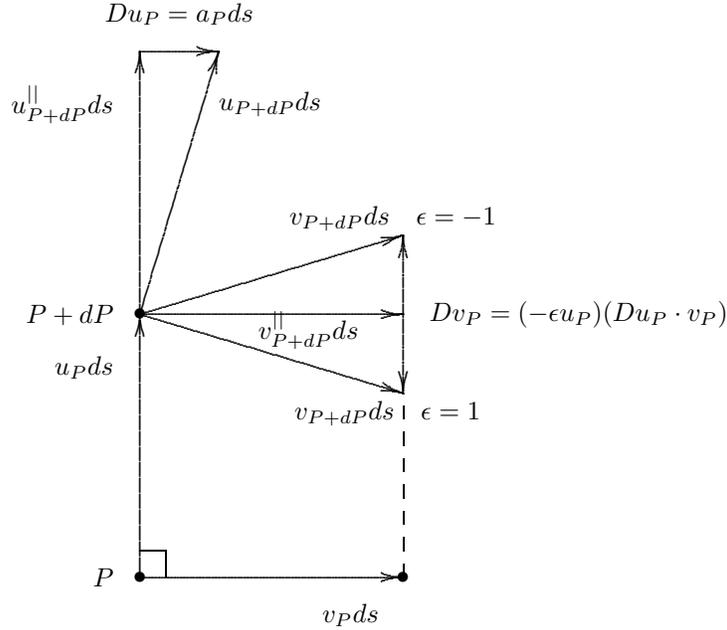

\typeout{figure 1}
$$ \vbox{
\beginpicture
\setcoordinatesystem units <3.5cm,3.5cm> point at  0 0  

\put {\mathput{\bullet}}                at  0 0
\put {\mathput{\bullet}}                at  0 1
\put {\mathput{\bullet}}                at  1 0

\put {\mathput{P}} [r]                at  -0.1 0
\put {\mathput{P+dP}} [r]             at  -0.1 1
\put {\mathput{u_P ds}} [r]           at  -0.1 0.8
\put {\mathput{u^{||}_{P+dP} ds}} [r] at  -0.1 1.8
\put {\mathput{u_{P+dP} ds}} [l] at  0.3 1.8
\put {\mathput{Du_P = a_P ds}} [b] at  0.15 2.1

\put {\mathput{v_P ds}} [t]           at  0.8 -0.1
\put {\mathput{Dv_P = (-\epsilon u_P)(Du_P \cdot v_P)}} [l] at 1.1 1
\put {\mathput{v^{||}_{P+dP} ds}} [l] at  0.45 .94

\put {\mathput{v_{P+dP} ds \quad \epsilon=-1}} [b] at  0.96 1.32
\put {\mathput{v_{P+dP} ds \quad \epsilon= 1}} [t] at  0.94 0.67


\arrow <.3cm>  [.1,.4]    from  0 0 to 0 1 
\arrow <.3cm>  [.1,.4]    from  0 0 to 1 0
\arrow <.3cm>  [.1,.4]    from  0 1 to 1 1
\arrow <.3cm>  [.1,.4]    from  0 1 to 0 2

\arrow <.3cm>  [.1,.4]    from  0 1 to 0.3 2
\arrow <.3cm>  [.1,.4]    from  0 2 to 0.3 2

\arrow <.3cm>  [.1,.4]    from  0 1 to 1 0.7
\arrow <.3cm>  [.1,.4]    from  0 1 to 1 1.3

\arrow <.3cm>  [.1,.4]    from  1 1 to 1 1.3
\arrow <.3cm>  [.1,.4]    from  1 1 to 1 0.7

\putrule from 0.1 0 to 0.1 0.1
\putrule from 0 0.1  to 0.1 0.1

\setdashes
\putrule from 1 0 to 1 0.7

\endpicture}$$
\caption{The Fermi transport argument. The unit vector $v$ orthogonal to $u$ must undergo a rotation ($\epsilon=1$) or boost ($\epsilon=-1$) in order to remain
orthogonal to $u$ as $u$ itself undergoes this same motion relative to the parallel transported direction $u^{||}$. Only the component of $v$ along the direction in which the tip of $u$ is moving must change, along the direction $u$ by an amount $Du\cdot v$.}
\end{figure}

Here $u^{(||)}$ and $v^{(||)}$ denote the parallel transported vectors from the point $P$ to the nearby point $P+dP$ on the curve. The unit vector $u$ has (pseudo-)rotated from $u^{(||)}$ by the amount $Du = (u-u^{(||)})ds = a ds$, which has the (pseudo-)angle as its magnitude.
In order for $v$ to remain orthogonal to $u$, 
it must undergo the same (pseudo-)rotation (boost/rotation) in the plane of $u$ and $a$. The difference vector $Dv = (v-v^{(||)})ds$ must then have the direction $-\epsilon u$ shown in the figure, but since only the component of $v$ in the plane of $u$ and $a$ need change, 
the scalar amount must be the projection of $v$ along $Du = a ds$, namely 
$v\cdot a ds$,
so 
$Dv = (-\epsilon u) (v\cdot a ds)$. Dividing through gives the Fermi relation
$Dv/ds = -\epsilon u \, a\cdot v$, representing the minimal (pesudo-)rotation necessary to keep $v$ orthogonal to $u$.

Next Fermi introduces coordinates in the following way. Complete $u$ to an orthonormal frame along the curve by adding the vectors $e_i$, $i=1,\ldots, n-1$, with $n=4$ in the case of a spacetime, where the orthogonal vectors are Fermi transported. For each unit vector $n=n^i e_i$ in the tangent space of a point on the curve at $x^\alpha(s)$, send out a geodesic in its direction and assign coordinates $(s,y^i)$ to points along it, where $y^i =\tilde s n^i$ and $\tilde s$ is the arclength along the geodesic, nicely illustrated by Fig.~13.4 in Misner, Thorne and Wheeler\cite{mtw} for the more general case of any completion of $u$ to an orthonormal frame. Fig.~2 gives the simpler picture analogous to Fig.~1. In the spacetime context $\{e_\alpha\} = \{u,e_i\}$ is a locally nonrotating (the spatial axes are fixed with respect to gyroscopes) proper frame for an observer with this world line, naturally called a Fermi frame, adapted to the observer's local time direction and local rest space.

\begin{figure}
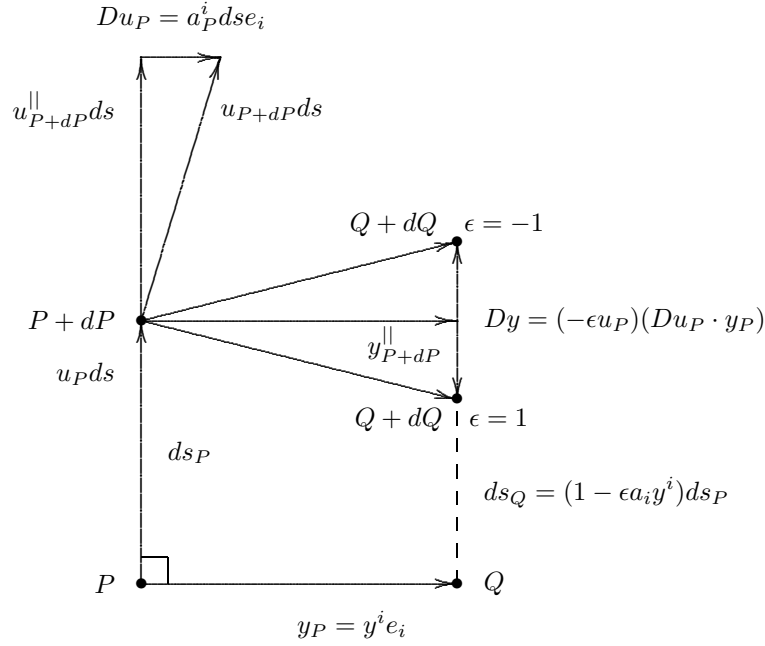

\typeout{figure 2}
$$ \vbox{
\beginpicture
\setcoordinatesystem units <3.5cm,3.5cm> point at  0 0  
\def\q{1.2} \def\qr{1.3} \def\qa{1.16} \def\qb{1.15}

\put {\mathput{\bullet}}                at  0 0
\put {\mathput{\bullet}}                at  0 1
\put {\mathput{\bullet}}                at  1.2 0
\put {\mathput{\bullet}}                at  1.2 1.3
\put {\mathput{\bullet}}                at  1.2 0.7

\put {\mathput{P}} [r]                at  -0.1 0
\put {\mathput{P+dP}} [r]             at  -0.1 1
\put {\mathput{u_P ds}} [r]           at  -0.1 0.8
\put {\mathput{u^{||}_{P+dP} ds}} [r] at  -0.1 1.8
\put {\mathput{u_{P+dP} ds}} [l]      at  0.3 1.8
\put {\mathput{Du_P = a_P^i ds e_i }} [b] at  0.15 2.1
\put {\mathput{y^{||}_{P+dP}}} [t] at  1 .98

\put {\mathput{Q}} [l]                at  1.3 0
\put {\mathput{ds_P}} [l]             at  0.1 0.5
\put {\mathput{ds_Q = (1-\epsilon a_i y^i) ds_P}} [l]  at 1.3 0.35

\put {\mathput{y_P=y^i e_i}} [t]           at  0.8 -0.1
\put {\mathput{Dy = (-\epsilon u_P)(Du_P \cdot y_P)}} [l] at 1.3 1

\put {\mathput{Q+dQ \quad \epsilon=-1}} [b] at  1.16 1.32
\put {\mathput{Q+dQ \quad \epsilon= 1}} [t] at  1.14 0.67


\arrow <.3cm>  [.1,.4]    from  0 0 to 0 1 
\arrow <.3cm>  [.1,.4]    from  0 0 to 1.2 0
\arrow <.3cm>  [.1,.4]    from  0 1 to 1.2 1
\arrow <.3cm>  [.1,.4]    from  0 1 to 0 2

\arrow <.3cm>  [.1,.4]    from  0 1 to 0.3 2
\arrow <.3cm>  [.1,.4]    from  0 2 to 0.3 2

\arrow <.3cm>  [.1,.4]    from  0 1 to 1.2 0.7
\arrow <.3cm>  [.1,.4]    from  0 1 to 1.2 1.3

\arrow <.3cm>  [.1,.4]    from  1.2 1 to 1.2 1.3
\arrow <.3cm>  [.1,.4]    from  1.2 1 to 1.2 0.7

\putrule from 0.1 0 to 0.1 0.1
\putrule from 0 0.1  to 0.1 0.1

\setdashes
\putrule from 1.2 0 to 1.2 0.7

\endpicture}$$

\caption{The Fermi normal coordinate first order metric derivation. To first order the coordinates $y^i$ remain orthonormal and orthogonal to $s$, but the (pseudo-)rotation of $u$ relative to the parallel transported direction $u^{||}$ causes the arclength along a nearby curve of constant (small) $y^i$ (the dashed line) to stretch/shrink compared to the arclength $s$
($ds_Q$ versus $ds_P$).
}
\end{figure}

Along the curve itself, these are orthonormal coordinates. As one moves slightly off the curve, the effect of the (pseudo-)rotation of $u$ on a nearby curve of points all sharing the same coordinates $y^i$ causes the arclength $ds_Q$ separating the points $Q$ and $Q+dQ$ to shrink/stretch relative to the arclength $ds_P$ separating corresponding points $P$ and $P+dP$ on the original curve. The change $Du_P = a^i ds\, e_i$ is a (pseudo-)angle times a unit vector, but only the component of $y_P=y^i e_i$ along this unit vector changes by this (pseudo-)angle times its length to get the arclength of the change. Thus the dot product 
$Du \cdot y_P = a^i ds_P e_i \cdot y^j e_j = a_i y^i ds_P$ gives the amount by which $ds_Q$ differs from $ds_Q$, with sign $\epsilon$ as in the figure: $ds_Q=(1-\epsilon a_i y^i) ds_P$. To first order the square is just $ds_Q{}^2 = (1-2\epsilon a_i y^i)ds_P{}^2$. To first order only the metric component along the curve changes
\beq
  ``ds^2" = ds_Q{}^2 +\eta_{ij} dy^i dy^j +O(y^2)
       = \epsilon(1-2\epsilon a_i y^i)ds_P{}^2 +\eta_{ij} dy^i dy^j  +O(y^2)\ ,
\eeq
(quotes to distinguish the line element symbol from the arclength differential along the curve, $\eta_{ij}$ for the flat orthonormal coordinate 3-metric of the appropriate signature) or dropping the subscript:
\beq
  ``ds^2" 
       = \epsilon(1-2\epsilon a_i y^i)ds^2 +\eta_{ij} dy^i dy^j  +O(y^2)\ .
\eeq

This is most useful for a timelike curve in spacetime, where $s=\tau$ is the proper time along the world line and this becomes
\beq
  ds^2 
       = -(1+2 a_i y^i)d\tau^2 +\delta_{ij} dy^i dy^j  +O(y^2)\ ,
\eeq
which is Eq.~(13.71) of Misner, Thorne and Wheeler\cite{mtw} with zero Fermi rotation of the spatial axes. These are called `Fermi normal coordinates,' and represent the locally nonrotating (with respect to gyroscopes) `proper reference frame' of a test observer following this world line. Along a geodesic, the metric is then the usual flat one up to first order in the flat spatial coordinates, characterized by the fact that along the curve itself in these coordinates, the connection components vanish and the metric  components are those of flat spacetime in inertial coordinates. 

These were first used by Levi-Civita to study geodesic deviation in 1926,\cite{levicivita26} as discussed by Manasse and Misner,\cite{manasse} who construct the Fermi frame and compute the Fermi normal coordinate metric up to the second order terms where the curvature tensor along the world line appears as in Riemann normal coordinates (hence the name `Fermi normal coordinates') and evaluate it for a radial geodesic in the Schwarzschild metric. O'Raifeartaigh\cite{raifeartaigh} investigated how one could duplicate the conditions of locally flat metric and vanishing connection components along an arbitrary accelerated curve, 
following up earlier work by Schouten.\cite{schouten35,schouten}
Synge discusses Fermi-Walker transport and Fermi coordinates at length in his 1960 book on general relativity.\cite{synge}

Ironically Levi-Civita is the original source claiming that Fermi established that one could find local coordinates along an arbitrary curve for which the connection components vanish and even gives a general argument why this should be so (footnote on p.~167 of ref.~\citen{levibook}). This claim is repeated by Schouten, Misner, Pirani\cite{pirani1} and others,\cite{marzlin} but Fermi only shows this for a geodesic, not an arbitrary curve, a puzzling fact. In Levi-Civita's discussion of geodesic deviation,\cite{levicivita26} he makes this general claim about arbitrary curves, but then only explicitly constructs Fermi coordinates for a geodesic. However, the resolution of the puzzle is that for nongeodesics, one must give up Fermi-Walker transport in the construction, using only parallel transport, as explicitly described by the 1-dimensional submanifold specialization of the more general discussion of O'Raifeartaigh.\cite{raifeartaigh} This breaks the link between the adapted coordinates along the world line and the nice orthogonal decomposition of the tangent space associated with the observer's local splitting of space and time, making the construction uninteresting from the point of view of gravitational theory.

Ni and Zimmerman\cite{nizim} followed up the Misner, Thorne and Wheeler discussion\cite{mtw} to evaluate the metric up to second order (curvature terms) for the more general rotating Fermi coordinate system along an arbitrary world line, the logical conclusion of the calculations started by Fermi and Walker. 
Their result shows how the gravitoelectric and gravitomagnetic contributions at the connection level due to the acceleration of the observer and the rotation of the observer spatial frame (the observer's GE and GM fields) and the gravitoelectric and gravitomagnetic parts of the curvature tensor measured by the observer all contribute to the description of the local proper frame of the observer nearby its world line.
The metric takes the form
\begin{eqnarray}
ds^2 &=& -[1+2 a\cdot x +(a\cdot x)^2 - (\omega\times x)^2
          + R_{0ioj} x^i x^j] dx^0{}^2\nonumber\\
      && + 2[ (\omega\times x)_i +{\textstyle\frac23} R_{0lmi}x^l x^m] dx^0 dx^i\nonumber\\
      && + [\delta_{ij} - {\textstyle\frac13} R_{iljm}x^l x^m] dx^i dx^j
         + O(x^3)\ ,
\end{eqnarray}
where the components of the acceleration $a_i$ of the world line and angular velocity $\omega_i$ of the orthonormal spatial frame $e_i$ 
(relative to a Fermi-transported spatial frame) are expressed in that frame, while the components of the Riemann tensor are evaluated in the orthonormal frame $e_0, e_i$ along the world line at $x^i=0$, where $e_0$ is the 4-velocity of the world line.

Even if one follows the Fermi normal coordinate construction exactly, one has problems with coordinate singularities and failure of the coordinates to cover all of spacetime even in the case of flat spacetime, as in the Rindler metric built on an observer family each member of which undergoes constant acceleration.
Marzlin\cite{marzlin} examines this difficulty with extending the coordinates away from the world line and suggests a modification of the construction to widen their extendibility. M\"arzke-Wheeler coordinates\cite{MW} are another option for correcting these difficulties, discussed more recently by Pauri and Vallisneri\cite{pauri} and 
Dolby and Gull,\cite{dolby} involving radar time.

Walker's approach to Fermi's discussion is based on Riemann normal coordinates at two nearby points along the timelike or spacelike curve in an $n$-dimensional space, which he uses to compute to second order in the remaining Fermi coordinates $y^i$ the rate $ds_Q/ds_P$ at which arclength changes on a nearby curve traced out by a point Q as in the above discussion, namely $y^i=y^i_0$, $s$ varying. Along a timelike curve in a spacetime, this shows how the tidal curvature affects the proper time of clocks carried in the observer proper frame as their spatial distance from the observer world line increases enough to begin to detect the effects of the curvature of spacetime. This then enables him to find the energy of each point in a small test body with respect to a test observer at any fixed reference point in the body chosen as the given world line, which for a small rigid body reproduces an earlier result, the apparent goal of his investigation.

Walker starts by introducing an arbitrary orthonormal frame $\{e_\alpha\}$ defined along the curve and the corresponding components of the connection induced along the curve
\beq
     \frac{D e_\alpha}{ds} = e_\beta W^\beta{}_\alpha\ ,
\eeq
which defines a mixed tensor whose totally contravariant form $W^\sharp$ or covariant form $W^\flat$ is antisymmetric, as follows from differentiating $e_\alpha\cdot e_\beta = \eta_{\alpha\beta}$
\beq
   W_{(\alpha\beta)} 
= \frac{D e_{(\alpha}}{ds}  
  \raise 7pt
\hbox{$\cdot\ e_{\beta)}$}  = 0\ .
\eeq
With hindsight we know that an antisymmetric second rank tensor can be expressed in terms of its electric and magnetic parts with respect to a unit vector direction
\beq
 W^{\alpha\beta} = [u\wedge A]^{\alpha\beta} + B^{\alpha\beta} \ ,
\eeq
where $A$ and $B$ are a vector and second rank antisymmetric tensor both orthogonal to $u$.

If one desires $u=u^\alpha e_\alpha$ to have constant components in such a frame along the curve, i.e. the frame vectors maintain constant angles with respect to the tangent $u$, then
\beq
   a^\alpha = \frac{D u^\alpha}{ds} = W^\alpha{}_\beta u^\beta = A^\alpha\ ,
\eeq
but this leaves $B$ arbitrary, describing the rotation of the frame about the direction $u$. For a timelike curve in spacetime, if one chooses a frame containing $u$, then $B$ is the angular velocity of the remaining spatial frame vectors in the local rest space of $u$, with respect to gyro-fixed axes, also called the Fermi rotation of the frame.

Without saying this, Walker notes that the simplest choice of $W$ amounts to setting $B=0$, which corresponds to Fermi-Walker transport of the frame along the curve. For a geodesic, he notes that it is natural to pick the orthonormal frame $\{e_\alpha\}$ to contain its tangent $u$, and for an accelerated curve, the Frenet-Serret frame (containing $u$) is suggested, in order to construct the family of Riemann normal coordinates used in his calculations (one coordinate system for each point on the given curve, which in general can agree with a Fermi coordinate system only at that given point). In studying nearby curves, he chooses an orthonormal frame containing $u$ whose remaining frame vectors are Fermi-Walker transported along the given curve. It is this frame that is used to evaluate the energies associated with the world lines of nearby curves in the case of a timelike curve in spacetime, as seen by a test observer moving along that curve.
(Ni and Zimmerman\cite{nizim} also give the coordinate accelerations of the world lines of these points, generalized to include rotation.)

\section{Why Useful?}

Fermi-Walker transport is useful because it describes the behavior of the spin vector $S$ of a torque-free test gyroscope carried by a test observer along its world line with 4-velocity $u$ (see Misner, Thorne and Wheeler\cite{mtw})
\beq
   u\cdot S = 0\ , \qquad
\frac{D_{\rm(fw)}S}{d\tau} = 0\ .
\eeq
The spatial vectors of a Fermi frame along the world line are therefore locally nonrotating with respect to a set of three independently oriented test gyros carried by the test observer and span the associated local rest space. Along a test particle in free motion along a geodesic, the Fermi frame is parallel transported along the world line and gives a tool for operationally measuring tidal effects of spacetime curvature along the world line.
However, in black hole spacetimes and the larger family of stationary axisymmetric spacetimes containing them, many families of accelerated test observers in uniform circular motion are defined by various aspects of the spacetime geometry and symmetry, so accelerated curves are important in interpreting this geometry and here Fermi-Walker transport is essential. Thus one is interested in the Fermi frame along both geodesics and accelerated curves.

\section{Geodesics in Black Hole Spacetimes}

For a proper time parametrized timelike geodesic world line of a test particle with 
4-velocity $u^\alpha =dx^\alpha/d\tau$, 4-momentum
$P^\alpha =\mu u^\alpha$, and vanishing acceleration $a^\alpha = Du^\alpha/d\tau =0$,
the rest mass $\mu$ provides one constant of the motion $P \cdot P = -\mu^2$. For black hole spacetimes (and in fact the entire Carter family of type D solutions),
the two Killing vectors $\xi[t]$, $\xi[\phi]$ associated with stationary axisymmetry together with the existence of a symmetric Killing tensor $K_{\alpha\beta}$ lead to three additional constants of the motion: $E=-\xi[t] \cdot P$ (energy at infinity) and $L_z = \xi[\phi] \cdot P$ (axial component of angular momentum) and either $\mathcal{Q}$ or $\mathcal{K}$, quadratic constants of the motion associated with the Killing tensor.\cite{carter} These allow the second order geodesic equations to be reduced to a first order system of four differential equations which can be interpreted in terms of motion in a potential for each of the Boyer-Lindquist coordinate variables. Marck,\cite{marck1,marck2,marck3} with some initial guidance from Carter, later showed that the Killing tensor also allowed one to reduce the construction of a Fermi frame along a geodesic to a single first order differential equation for a rotation angle. 

The tangent to an affinely parametrized timelike geodesic is parallel transported along the geodesic. If one can come up with a second independent parallel transported vector, its spatial projection will also be parallel transported and by normalization, one has the first two vectors of a Fermi frame, leaving the second two in the orthogonal 2-plane defined up to the angle of rotation in that plane.

For a Killing vector $\xi$, the quantity $\xi\cdot u$ is a conserved quantity (specific energy or angular momentum in our case) along a geodesic
\beq
  \xi_{(\alpha;\beta)} =0= a^\alpha\ ,\qquad
 \frac{D}{d\tau}(\xi_\gamma u^\gamma) 
= \xi_\alpha a^\alpha + \xi_{(\alpha;\beta)} u^\beta u^\alpha = 0\ ,
\eeq
using the chain rule $DX/\tau =u\cdot \nabla X$ for differentiating a field $X$ along the curve. 

For a symmetric second rank tensor which is a Killing tensor
\beq
  K_{[\alpha\beta]}=0\ ,\quad
   K_{(\alpha\beta;\gamma)}=0\ , 
\eeq 
one can define a vector
$W^\alpha = K^\alpha{}_\beta u^\beta$, a scalar $Q=K_{\alpha\beta}u^\alpha u^\beta 
= W_\alpha u^\alpha$ and another vector
$[P(u)W]^\alpha = (\delta^\alpha{}_\beta +u^\alpha u_\beta) W^\beta
=W^\alpha +Q u^\alpha$ which is orthogonal to $u$. Thus
$W = P(u)W - Q u$ is the orthogonal decomposition of $W$ with respect to $u$.
Similar calculations show that $Q$ is a quadratic constant of the motion, and the intrinsic derivatives of $W$ and $P(u)W$ along $u$ are both orthogonal to $u$, but in general nonzero. This fails to lead to a second parallel-transported direction.

However, black hole spacetimes also have a second rank Killing-Yano tensor, which is just an antisymmetric tensor satisfying
\beq
  F_{(\alpha\beta)}=0\ ,\qquad
  F_{\alpha(\beta;\gamma)}=0\ , 
\eeq
found first by Penrose and Floyd.\cite{penrose}
Its square $F^\alpha{}_\gamma F^\gamma{}_\beta$ is automatically a (symmetric) Killing tensor, which is just $K$ itself in this case. This has the advantage that the vector $v^\alpha = F^\alpha{}_\beta u^\beta$ is not only parallel transported along this geodesic but is automatically orthogonal to $u$ (since $v^\alpha u_\alpha =F_{\alpha\beta}u^\alpha u^\beta=0$), and so only has to be normalized (provided that it is not zero) to get a second Fermi frame vector.
These may be completed to an orthonormal frame by adding two spacelike unit vectors defined up to an angle of rotation in the plane orthogonal to the first two vectors. The equations of parallel transport leads to a first order differential equation for this angle to fix those vectors to also be parallel transported, leading to the Fermi frame. Because of the constants of the motion, this equation may be solved by an integral formula involving those constants.

Marck extends the Fermi frame calculation to null geodesics in his second article and then to spacetimes with two Killing vectors and a Killing-Yano tensor in his third article. He then uses the results to study tidal curvature effects along geodesics in nonrotating black hole spacetimes.\cite{marck4}

\section{Circular Orbits in Stationary Axisymmetric Spacetimes}

Accelerated orbits are much more difficult to treat, so one must assume more symmetry in the orbits themselves to make progress, and one must distinguish between parallel transport and Fermi-Walker transport. Circular orbits in stationary axisymmetric spacetimes (especially black hole spacetimes) are very interesting for many reasons, and turn out to have nice geometry associated with these transports which can be explicitly described without having to integrate any remaining differential equations. Although one can state the final result which solves the equations of parallel transport or Fermi-Walker transport for any vector along such orbits,\cite{bjm2002} one can build up the same formula by including a number of relativistic effects one by one, leading to the final factored form of that formula. This gives a nice geometric interpretation to the result.

Consider general accelerated but constant speed circular orbits in the equatorial plane $\theta=\pi/2$ of a black hole in Boyer-Lindquist coordinates $t,r,\theta,\phi$, with mass and specific angular momentum parameters $m$ and $0\leq a<1$. This `plane' admits a pair of oppositely rotating periodically intersecting circular geodesics which are timelike outside a certain radius and allow the study of various so called `clock effects' by comparing either observer or geodesic proper time periods of orbital circuits defined by the observer or the geodesic crossing points. This can be extended to a comparison of parallel transported vectors, corresponding to special holonomy transformations, and with some modifications to Fermi-Walker transport corresponding to a sort of `spin holonomy.' 

The stationary circular orbits containing every second meeting point of the oppositely rotating circular geodesics departing from an initial  crossing point are called geodesic meeting point observers (GMPOs).\cite{bjm2002}
Since they are intimately connected to the properties of parallel transport by their definition as having a parallel transported tangent vector, it should be no surprise that the circular geodesics (and these GMPOs) play a key role in the interpretation of the parallel transport transformation along a general circular orbit. 
If $\zeta_{\rm(geo)-}<0<\zeta_{\rm(geo)+}$ are the angular velocities ($d\phi/dt$) of oppositely rotating geodesics, the GMPOs have their average angular velocity
$\zeta_{\rm(gmp)} = \frac12(\zeta_{\rm(geo)+} + \zeta_{\rm(geo)-})$, which is nonzero only for a rotating black hole where asymmetry exists between the corotating $(+)$ and counterrotating $(-)$ geodesics.

\section{Closed $\phi$ loops}

First we study the various relativistic effects which describe parallel transport around accelerated circular orbits and then for Fermi-Walker transport around the same orbits. The primary effect to consider arises from the geometry of a single $\phi$ coordinate loop at fixed time $t$, so that the tangent vector to the orbit lies in the local rest space of the ZAMOs (zero angular momentum observers), also called the locally nonrotating observers, whose 4-velocity is the unit normal $n$ to the $t$ hypersurfaces. Let $R=g_{\phi\phi}^{1/2}$ be the circumferential radius of the orbit at coordinate radius $r$, and let $\rho = |-(\ln R)_{,r}/g_{rr}^{1/2}|^{-1}$ be its Lie relative curvature,\cite{idcf1} or more descriptively, the ZAMO intrinsic radius of turning.

\begin{figure}
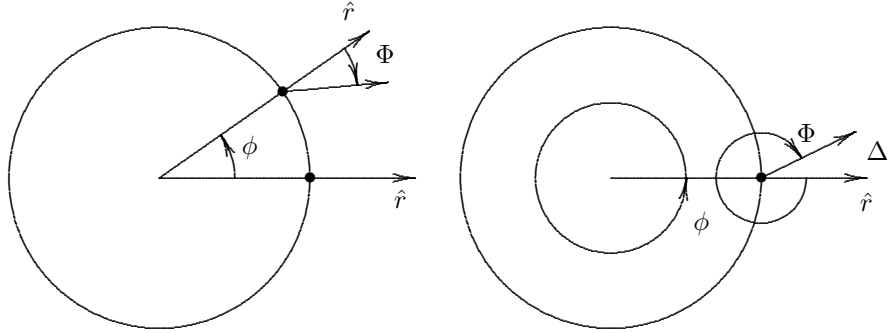

\typeout{figure 3}
$$ \kern3cm \vbox{
\beginpicture
\setcoordinatesystem units <2cm,2cm> point at 0 0   

\put {\mathput{\bullet}}                at  1 0
\put {\mathput{\phi}}                at  .6 -.3
\put {\mathput{\Phi}}                at 1.3 .3
\put {\mathput{\hat r}} [t]             at  1.7 -0.1
\put {\mathput{\Delta}} [l]        at  1.7 0.18

\circulararc 360 degrees from 1 0 center at 0 0
\circulararc 360 degrees from .5 0 center at 0 0
\circulararc -334 degrees from 1.3 0 center at 1 0

\arrow <.3cm>  [.1,.4]    from  0 0 to 1.7 0 
\arrow <.3cm>  [.1,.4]    from  .489 -.1 to .505 0
\arrow <.3cm>  [.1,.4]    from  1.19 .2315 to 1.2696 .1315
\arrow <.3cm>  [.1,.4]    from  1 0 to 1.6292 .3068

\setcoordinatesystem units <2cm,2cm> point at 3 0   

\put {\mathput{\bullet}}                at  1 0
\put {\mathput{\bullet}}                at  .819 .574
\put {\mathput{\phi}}                at  .6 .2
\put {\mathput{\Phi}}                at 1.5 .825
\put {\mathput{\hat r}} [t]             at  1.6 -0.08
\put {\mathput{\hat r}} [br]             at  1.3 1.05

\circulararc 360 degrees from 1 0 center at 0 0
\circulararc 35 degrees from .5 0 center at 0 0
\circulararc -30 degrees from 1.229 .8604 center at .8192 .5736

\arrow <.3cm>  [.1,.4]    from  0 0 to 1.7 0 
\arrow <.3cm>  [.1,.4]    from  0 0 to 1.393 .9751
\arrow <.3cm>  [.1,.4]    from  .8192 .5736 to 1.5165 .63461
\arrow <.3cm>  [.1,.4]    from  0.458 .1868 to 0.4096 .2868
\arrow <.3cm>  [.1,.4]    from  1.296 .71718 to 1.3173 .61718

\endpicture}$$
\vglue1.5cm
\caption{Parallel transport around a circular orbit in its own `plane'. $d\Phi/d\phi=1$ in a flat geometry, but curvature causes this relative frequency to differ from 1. When $0<d\Phi/d\phi<1$ as in this diagram, the transported direction advances in the orbital direction because the parallel transport angle $\Phi$ lags behind the orbital angle $\phi$.
}\label{fig:circles}
\end{figure}

For a tangent vector in the $t$-$\phi$ subspace of the tangent space like the initial radial direction in Fig.~\ref{fig:circles}, parallel transport around an orbital interval of angle $\phi$ leads to a compensating (oppositely directed) angle $\Phi$ of the parallel transported direction relative to the actual radial direction which cancel each other out in a flat geometry: $d\Phi/d\phi=1$.
When the intrinsic geometry of this plane is not flat, after completing one circuit of the orbit, the radial direction will be rotated forward by the angle $\phi-\Phi$, an advance described by the relative frequency rate
$1-d\Phi/d\phi$. The amount is nicely interpreted by an explanation given by Thorne\cite{tho81,idcf2} using the tangent cone to the surface of revolution embedding diagram for the intrinsic geometry of the equatorial plane.

\begin{figure}
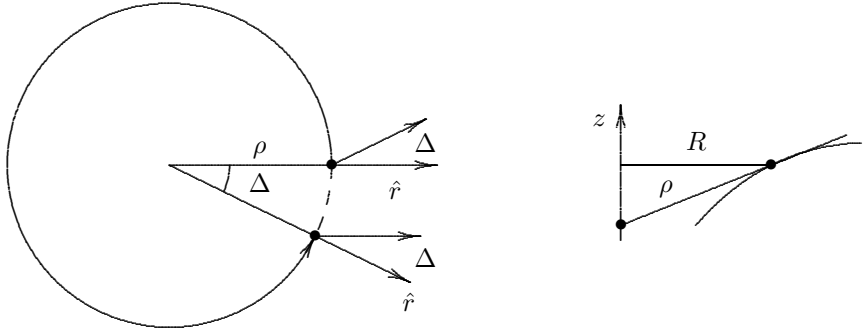

\typeout{figure 4}
\vglue 1.5cm
$$ \kern3cm \vbox{
\beginpicture
\setcoordinatesystem units <2cm,2cm> point at 0 0   
\put {\mathput{\bullet}}                at  1 0
\put {\mathput{\bullet}}                at  0 -0.4022
\put {\mathput{z}}     [r]           at  -0.1 0.3
\put {\mathput{R}}     [b]           at  0.5 0.1
\put {\mathput{\rho}}              at  0.3 -0.155

\putrule from 0 0 to 1 0
\plot 0 -0.4   1.5 0.2 /
\arrow <.3cm>  [.1,.4]    from  0 -0.5 to 0 0.4 

\setquadratic
\plot 0.5 -0.4  1 0  1.6 0.15 /
\setlinear

\setcoordinatesystem units <2cm,2cm> point at 3 0   

\put {\mathput{\bullet}}                at  1.078 0
\put {\mathput{\bullet}}                at  .9689 -.4725

\put {\mathput{\rho}}    [b]            at  .6 .05
\put {\mathput{\Delta}}                 at  .6 -.12
\put {\mathput{\hat r}} [t]             at  1.5 -0.1
\put {\mathput{\hat r}} [tl]             at  1.55 -.85

\circulararc 334 degrees from 1.0778 0 center at 0 0
\circulararc -26 degrees from .4 0 center at 0 0

\setdashes
\circulararc -26 degrees from 1.0778 0 center at 0 0
\setsolid

\arrow <.3cm>  [.1,.4]    from  1.078 0 to 1.7072 .3068
\arrow <.3cm>  [.1,.4]    from  0 0 to 1.778 0
\put {\mathput{\Delta}}                 at  1.7 .15

\arrow <.3cm>  [.1,.4]    from .9689 -.4725 to 1.6689 -.4725
\arrow <.3cm>  [.1,.4]    from  0 0 to 1.598 -.7793
\put {\mathput{\Delta}}                 at  1.7 -.62

\arrow <.3cm>  [.1,.4]    from  .91 -.5725 to .9689 -.4725

\endpicture}$$
\vglue0.5cm
\caption{The equatorial plane embedding argument. A piece of the  cross-section of the embedding surface of revolution is shown together with the tangent cone on the right. The orbit has proper circumference $2\pi R$.
Opening the flat tangent cone leads to the circle of radius $\rho$ on the left, 
where the two bullet points are identified,
and having a deficit angle $\Delta$ satisfying (solid arc plus dashed arc equals total circumference)
$2\pi R + \Delta \rho = 2\pi \rho$ or $\Delta = 2\pi (1 - R/\rho)$.
Parallel transport keeps the initial radial direction horizontal, so when it finishes its circuit, it has advanced by the deficit angle with respect to the final radial direction. Thus the deficit angle $\Delta$ is
exactly the total parallel transport angle for a complete orbital circuit, in the same direction as the orbit.
}\label{fig:cone}
\end{figure}

When the ratio $R/\rho<1$, one can imbed this geometry in Euclidean 3-space as a surface of revolution about the $z$-axis, where $R$ is the distance of a point on this surface from the $z$-axis and $\rho$ is the distance to the vertex of the tangent cone to the surface which lies on the $z$-axis as shown in Fig.~\ref{fig:cone}. This shows the relationship between the net parallel transport angle per completed orbital circuit
\beq
  \Delta = 2\pi (1 - R/\rho) > 0\ ,
\eeq
advancing in the same direction as the orbital direction. Since the $r,\theta,\phi$ coordinates are orthogonal and $\phi$ is a Killing coordinate, parallel transport does not affect the component of a spatial vector (in the local rest space of $n$) perpendicular to the equatorial plane, so this rotation angle in the $r$-$\phi$ tangent subspace completely describes the parallel transport around the closed $\phi$ loops in the intrinsic curved geometry of the plane. This corresponds to a relative frequency rate
$0< d\Phi/d\phi = R/\rho < 1$.

In fact one can express this in 3-dimensional matrix notation in these coordinates as follows. If $X(0)$ is the initial (component) tangent vector and $X(\phi)$ the final parallel transported vector after a change of orbital angle $\phi$, one has
\beq\label{eq:Aint}
   X(\phi) = e^{\phi A_{\rm(int)}} X(0)\ ,
\qquad
   [A_{\rm(int)}]^i{}_j = \frac{R}{\rho} [e_{\hat r}^\flat \wedge e_{\hat\phi}^\flat ]^i{}_j \ , 
\eeq
where $e^\flat_{\hat r} = g_{rr}^{1/2} dr$ 
and $e_{\hat\phi}^\flat = R d\phi$
are the orthonormal 1-forms, and the matrix $([A_{\rm(int)}]^i{}_j)$ consists of the coordinate components of a mixed second rank tensor, which is antisymmetric upon lowering its indices. This exponential representation of the parallel transport transformation arises as the solution of the constant coefficient linear system of intrinsic parallel transport equations for the coordinate components along the $\phi$-parametrized curves
\beq
   \frac{dX(\phi)^i}{d\phi} 
= A_{\rm(int)}{}^i{}_j X(\phi)^j\ , 
\eeq
which corresponds to vanishing intrinsic covariant derivative of $X$ along $\phi$
\beq
        \frac{D_{\rm(int)}X(\phi)^i}{d\phi} 
  = \frac{dX(\phi)^i}{d\phi} 
       -A_{\rm(int)}{}^i{}_j X(\phi)^j =0\ .
\eeq
By applying this to the vector $e_{\hat r}^\alpha$, 
which has the covariant derivative $-A^i{}_j e_{\hat r}^j$,
one sees that
\beq
  \frac{D_{\rm(int)} e_{\hat r}}{d\phi} = \frac{R}{\rho} e_{\hat\phi}\ ,
\eeq
which in turn allows the contravariant form of the tensor $A_{\rm(int)}$ to be re-expressed as
\beq
    A_{\rm(int)}^\sharp = e_{\hat r}\wedge  \frac{D_{\rm(int)} e_{\hat r}}{d\phi}\ .
\eeq

However, the spacetime parallel transport issue is distinct from the intrinsic geometry parallel transport, since additional extrinsic curvature terms enter the calculation. It turns out (hindsight) that the radial variation of the tilting of the Killing $t$-coordinate lines (static observer world lines) away from the normal direction causes the 2-plane of this intrinsic parallel transport rotation to tilt as well in the context of spacetime parallel transport (from the extra Christoffel symbol terms due to the extrinsic curvature). This variation can be described by the angular velocity 
$\zeta_{\rm(gmp)}=-g_{t\phi,r}/g_{\phi\phi,r}$ of the GMPOs,\cite{bjm2002} and in the spacetime context, evaluating the parallel transport equations explicitly shows that in the above formula, one adds an additional term to the coefficient matrix which converts $e_{\hat\phi}^\flat= R d\phi$ into 
\beq\label{eq:Ygmp}
  Y(\zeta_{\rm(gmp)})^\flat = R (d\phi - \zeta_{\rm(gmp)}dt)
  = \gamma(\zeta_{\rm(gmp)})^{-1} e_{\hat\phi}(\zeta_{\rm(gmp)})^\flat
\ , 
\eeq
which is in the $\phi$ angular direction within the local rest space of the GMPOs, but is the Lorentz contraction of the original unit vector $e_{\hat\phi}$ in the local rest space of the ZAMOs in our intrinsic discussion by the relative gamma factor of the geodesic meeting point observers with respect to the ZAMOs.
Since the rotation plane belongs to the local rest space of the GMPOs, their 4-velocity $u(\zeta_{\rm(gmp)})$ is invariant under parallel transport along these orbits. 

The spacetime result is then
\beq\label{eq:A}
   X(\phi) = e^{\phi A} X(0)\ ,\qquad
   A^\alpha{}_\beta =-\Gamma^\alpha{}_{\phi\beta} 
        = \gamma(\zeta_{\rm(gmp)})^{-1}\frac{R}{\rho} 
         [e_{\hat r}^\flat \wedge 
               e_{\hat\phi}(\zeta_{\rm(gmp)})^\flat ]^\alpha{}_\beta
    \ , 
\eeq
which is the exponential solution of the linear system
\beq
   \frac{dX(\phi)^\alpha}{d\phi} 
= A{}^\alpha{}_\beta X(\phi)^\beta\ , 
\eeq
so that the parallel transport angle $\Phi$ in the 
$e_{\hat r}$-$e_{\hat\phi}(\zeta_{\rm(gmp)})$ plane satisfies 
\beq
    \frac{d\Phi}{d\phi} = \frac{\gamma(\zeta_{\rm(gmp)})^{-1} R}{\rho}
\ .
\eeq
This is the ratio of the Lorentz contraction of the ZAMO intrinsic circumferential radius of the orbit to the local rest space of the GMPOs in which the rotation takes place with the ZAMO intrinsic radius of turning, which sort of seems reasonable. Notice also that
the gamma factor increases the slowing down of the parallel transport rotation compared to the orbital rotation of the intrinsic parallel transport alone and when $R/\rho<1$ (when the orbital `plane' can be embedded in
Euclidean 3-space), leads to a prograde rotation with angular velocity $1-d\Phi/d\phi>0$ relative to the orbital angle.

As above one has the analogous relation
\beq\label{eq:Der}
  \frac{D e_{\hat r}}{d\phi} = \frac{d\Phi}{d\phi}\,
 e_{\hat\phi}(\zeta_{\rm(gmp)})\ ,
\eeq
which can be used to re-express (\ref{eq:A}) as
\beq
  A^\sharp = e_{\hat r}\wedge \frac{De_{\hat r}}{d\phi} \ .
\eeq

\section{Helical $\phi$ loops: observer-dependent circuits}

\begin{figure}
\typeout{*** EPS figure 5}
\centerline{
\epsfbox{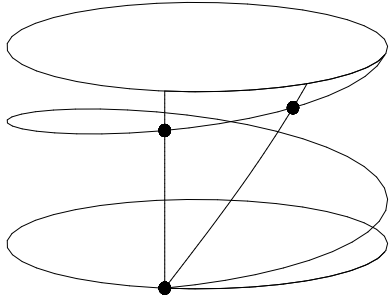}
}
\caption{The cylinder of circular orbits at a given radius. Once a circular orbit does not close, what constitutes one circuit of the hole depends on which stationary circularly rotating observer is the reference world line.}
\label{fig:helix}
\end{figure}

However, the closed $\phi$ loop orbits are spacelike curves with infinite angular velocity, and we are interested in timelike circular orbits, so we have to now tilt the orbits themselves in time. Moreover, once we `open the loop' by creating a helical curve in spacetime, another problem arises: how to define a complete circuit of the orbit.
In fact each stationary circularly rotating observer defines a complete circuit differently, so it must be kept in mind that this concept is clearly observer-dependent, as illustrated by Fig.~\ref{fig:helix}.

In a nonrotating (static) black hole, there is an obvious preferred choice of observer: the nonrotating static observers following the $t$-coordinate lines in the Boyer-Lindquist coordinates. The circuits are then simply characterized by $\Delta\phi=\pm 2\pi$, i.e. starting at a given $t$ line and then orbiting one loop around either direction to return to the same line. However, in a rotating black hole, many distinct choices exist all of which reduce to the preferred static observers in the limit of zero rotation:
\begin{itemize}
\item static observers
 or distantly nonrotating observers, following the time coordinate lines along the Killing vector field generating the stationary symmetry which reduces to time translation at spatial infinity,

\item ZAMOs or locally nonrotating observers, 
whose world lines are orthogonal to the time coordinate hypersurfaces,

\item geodesic meeting point observers, 
whose world lines contain every second meeting point of a pair of oppositely rotating geodesic orbits at a given radius,

\item extremely accelerated observers (EAOs), for which the magnitude of their acceleration is maximal (except near the hole, where it is minimal),

\item 
Carter observers,
whose 4-velocity is the along the intersection of the 2-plane of the 2 repeated null directions of the Riemann curvature tensor with the tangent 2-plane to the $t$-$\phi$ cylinder orbits of the stationary axisymmetric symmetry.

\end{itemize}

\begin{figure}
\typeout{*** EPS figure 6}
\centerline{
\epsfbox{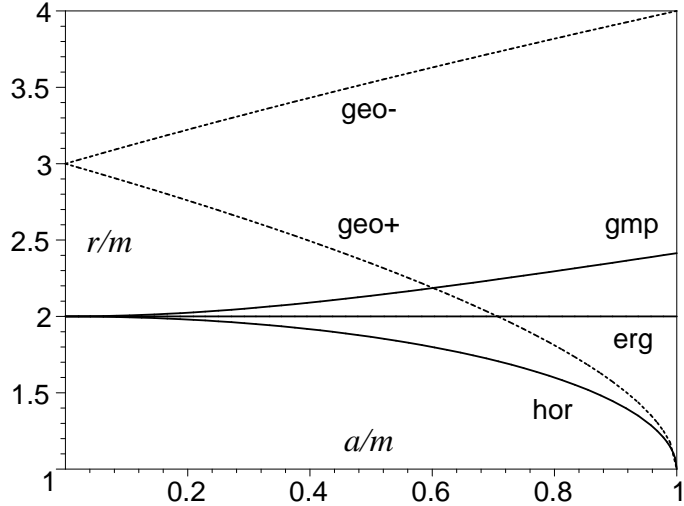}
}
\caption{The observer horizon radius for each of the geometrically defined stationary circularly rotating observers in the equatorial plane of a Kerr black hole for the physically interesting dimensionless angular momentum  parameter interval $0\leq a/m \leq 1$. The ZAMOs and Carter observers share the black hole horizon (hor) as their observer horizon, the static observers have the ergosphere boundary (erg) as their horizon, while the EAOs have their observer horizon at the radius where the counterrotating geodesics go null (geo$-$).}
\label{fig:horizons}
\end{figure}

Each of these observer families has the same limit at spatial infinity far from the hole, but approaching the hole, each encounters an observer horizon at which their 4-velocity goes null. Some continue to be defined by their geometrical properties as spacelike curves within this horizon, while others do not. For the equatorial plane of our present discussion, Fig.~\ref{fig:horizons} compares the various observer horizon radii, all of which lie in the interval $1\leq r/m \leq 4$.

Passing to the case of general stationary circularly rotating curves requires the further change in the tangent vector: 
$\partial_\phi \to \partial_\phi + \zeta^{-1} \partial_t$ and considering values of $ \zeta^{-1}$ away from zero, tilting the curves in spacetime with respect to the closed $\phi$ loops. This tangent vector corresponds to using the value of the coordinate $\phi$ along the curve as its parameter.
Switching to $t$ as a parameter along these curves leads to the rescaled tangent vector $\partial_t + \zeta \partial_\phi$, where $\zeta = d\phi/dt$ is the angular velocity of the curve. When nonnull 
($\Gamma(\zeta)^{-2} \neq0$), the tangent can be normalized to a unit vector with corresponding 1-form
\begin{eqnarray}
&&   u(\zeta) = \Gamma(\zeta) [\partial_t + \zeta \partial_\phi]\ ,\qquad
  u(\zeta)^\flat = \gamma(\zeta) R [d\phi - \bar\zeta dt]\ ,\nonumber\\
&& \Gamma(\zeta)^{-2} 
= -[\partial_t + \zeta \partial_\phi]\cdot 
           [\partial_t + \zeta \partial_\phi]\ ,
\end{eqnarray}
where $\bar\zeta$ is the angular velocity of the circular orbit with unit tangent 
$u(\bar\zeta)$ orthogonal to $u(\zeta)$ 
and $\gamma(\zeta)$ is the corresponding gamma factor with respect to the ZAMOs
\beq
u(\zeta) = \gamma(\zeta) [ n + \nu^{\hat\phi}(\zeta) e_{\hat\phi}]\ ,\ 
\gamma(\zeta) = [1- \nu^{\hat\phi}(\zeta)^2]^{-1/2}
= N \Gamma(\zeta)
\eeq
and $N=(-g^{tt})^{-1/2}$ is the lapse function. The relationship between the two relative velocities is a reciprocal one
\beq
  \nu^{\hat\phi}(\bar\zeta) = 1/\nu^{\hat\phi}(\zeta) =\bar\nu^{\hat\phi}(\zeta)\ ,
\eeq
where
\beq
   \nu^{\hat\phi}(\zeta) = \frac{R}{N}(\zeta + N^\phi)\ ,\qquad N^\phi = g_{\phi t}/g_{\phi\phi}
\eeq
relates the relative velocity to the angular velocity. 

The additional term in the tangent vector leads to an additional tilt in the angular direction of the plane of the parallel transport rotation as one moves away from $\zeta^{-1}=0$, corresponding to the angular direction of a new stationary circularly rotating observer. The previous factor picks up an additional term involving the angular velocity 
$\bar\zeta_{\rm(car)} = -g_{tt,r}/g_{t\phi,r}=a $
of the curves orthogonal to the Carter observers
\begin{eqnarray}
&&  Y(\zeta_{\rm(gmp)})^\flat 
= R (d\phi - \zeta_{\rm(gmp)}dt)
  = \gamma(\zeta_{\rm(gmp)})^{-1} e_{\hat\phi}(\zeta_{\rm(gmp)})^\flat
\nonumber\\
&&
\to 
  R[d\phi-\zeta_{\rm(gmp)} d t 
         -(\zeta_{\rm(gmp)}/\zeta)(d\phi-\bar\zeta_{\rm(car)} d t)]
\nonumber\\ 
\qquad
&& =\cases{
(1-\zeta_{\rm(gmp)}/\zeta) Y(\mathcal{Z}(\zeta))^\flat\ , 
        & $\zeta\neq \zeta_{\rm(gmp)}$\ ,\cr
 (\bar\zeta_{\rm(car)} - \zeta_{\rm(gmp)}) R d t\ , 
        & $\zeta = \zeta_{\rm(gmp)}$\ ,\cr 
 ( 1-\zeta_{\rm(gmp)}/\bar\zeta_{\rm(car)}) R d\phi\ , 
        & $\zeta = \bar\zeta_{\rm(car)}$\ ,\cr
}
\end{eqnarray}
where a map $\zeta \to \mathcal{Z}(\zeta)$
picking out the parallel transported direction is defined explicitly by
\beq\label{eq:Zzeta}
 \mathcal{Z}(\zeta) 
    = \zeta_{\rm(gmp)} 
   \frac{\zeta-\bar\zeta_{\rm(car)}}{\zeta-\zeta_{\rm(gmp)}}
\mathop{\longrightarrow}\limits^{g_{\phi t}\to0}
 -\frac{g_{tt,r}}{g_{\phi\phi,r}} \frac{1}{\zeta}
\ ,\eeq
and satisfies $\mathcal{Z}(\mathcal{Z}(\zeta))=\zeta$.
$\mathcal{Z}(\zeta)$ is the angular velocity of the stationary axisymmetric vector field tangent to the symmetry orbit which is covariant constant along the Killing trajectory with angular velocity $\zeta$, while
$Y(\mathcal{Z}(\zeta)) 
=  \gamma(\mathcal{Z}(\zeta))^{-1} e_{\hat\phi}(\mathcal{Z}(\zeta)) $ is the angular direction in the orthogonal subspace of the tangent space.
The limit $\lim_{\zeta^{-1}\to0} \mathcal{Z}(\zeta) =\zeta_{\rm(gmp)}$ returns us to the closed $\phi$ loop case.

Thus for $\zeta\neq\zeta_{\rm(gmp)}$
the parallel transport rotation takes place in the 2-plane spanned by
$e_{\hat r}\wedge e_{\hat\phi}(\mathcal{Z}(\zeta))$ and satisfies
\beq\label{eq:Phiphi}
     \frac{d\Phi}{d\phi} 
  = (1-\zeta_{\rm(gmp)}/\zeta) 
     \gamma(\mathcal{Z}(\zeta))^{-1} \frac{R}{\rho}\ .
\eeq
This corresponds to the exponential solution 
\beq\label{eq:Azeta}
   X(\phi) = e^{\phi A(\zeta)} X(0)\ ,\qquad
   A(\zeta)^\alpha{}_\beta 
    = -(\Gamma^\alpha{}_{\phi\beta}+\zeta^{-1}\Gamma^\alpha{}_{t\beta})
        = \frac{d\Phi}{d\phi}  
         [e_{\hat r}^\flat \wedge 
               e_{\hat\phi}(\mathcal{Z}(\zeta))^\flat ]^\alpha{}_\beta
\eeq
of the
parallel transport equations
\beq
   \frac{dX(\phi)^\alpha}{d\phi} 
= A(\zeta)^\alpha{}_\beta X(\phi)^\beta\ . 
\eeq
Note that the vector field $e_{\hat r}$, which is spatial with respect to all circularly rotating observers, satisfies
\beq\label{eq:Der2}
  \frac{D e_{\hat r}}{d\phi} = \frac{d\Phi}{d\phi}\,
 e_{\hat\phi}(\mathcal{Z}(\zeta))\ ,
\eeq
which again implies the relation
\beq
  A(\zeta)^\sharp = e_{\hat r}\wedge \frac{De_{\hat r}}{d\phi} \ .
\eeq

The $\phi$ parametrization of the circular orbit corresponds to defining circuits with respect to the $t$-coordinate lines. However, the additional factor $(1-\zeta_{\rm(gmp)}/\zeta)$ which now appears is exactly the factor needed to change the parametrization from the values of the angular coordinate $\phi$ along the orbit to the values of the new angular coordinate $\tilde\phi=\phi - \zeta_{\rm(gmp)}t$  dragged along by the GMPOs. Since $dt/d\phi=1/\zeta$ along these orbits, one immediately gets $d\tilde\phi/d\phi = (1-\zeta_{\rm(gmp)}/\zeta)$.
$\tilde\phi: 0\to\pm2\pi$ describes one complete circuit corotating or counterrotating with respect to the GMPOs. The frequency rate ratio between parallel transport angle and the GMPO angle is then directly analogous to the closed $\phi$ loop case
\beq
     \frac{d\Phi}{d\tilde\phi} 
  = \gamma(\mathcal{Z}(\zeta))^{-1} \frac{R}{\rho}\ .
\eeq

\begin{figure}
\typeout{*** EPS figure 7}
\centerline{
\epsfbox{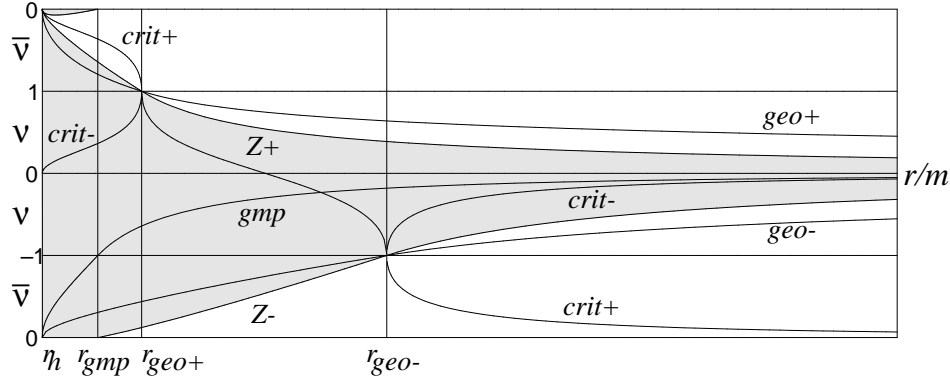}
}
\caption{The parallel transport boost and rotation velocity profile zones for all timelike, null or spacelike stationary circular orbits in the equatorial plane outside the black hole horizon at $r=r_h\approx 1.866m$ out to $r/m=6$, illustrated for $a/m=0.5$. 
The horizontal axis is the ZAMO velocity profile ($\nu^{\hat\phi}=0$); other observer horizons occur where $|\nu^{\hat\phi}|=1$. The upper and lower borders $\bar\nu{}^{\hat\phi}=0$ of the rectangular plot are identified and correspond to the closed $\phi$ orbits.
For spacelike orbits $|\nu^{\hat\phi}|>1$, the reciprocal velocity $\bar\nu^{\hat\phi}= 1/\nu^{\hat\phi}$ is plotted with decreasing absolute value as one moves away from the horizontal axis. The boost region is shaded, leaving the complementary unshaded region as the rotation zone, separated by the velocity profiles of $\nu^{\hat\phi}(\mathcal{Z}(\zeta_\pm))$, 
denoted by $Z\pm$,
corresponding to null rotations. The circular geodesic profiles cross these velocity curves at their corresponding mutual horizons. The velocity profile of the GMPOs cuts through the middle of the shaded boost region. The three vertical lines inside the plot indicate the GMPO horizon and the corotating ($geo+$) and counterrotating ($geo-$) circular geodesic horizons. Finally the critical velocities ($crit\pm$) of the magnitude of the acceleration  are shown, symmetric about $\nu^{\hat\phi}=1$ since they satisfy $\nu^{\hat\phi}_{\rm(crit)+} \nu^{\hat\phi}_{\rm(crit)-}=1$, 
with the $crit-$ curve corresponding to the extremely accelerated observer velocity.
For completeness, the spin critical observer\cite{idcf2} velocity profile which connects the two vertices where the $crit+$ and $crit-$ curves meet at the two radii $r_{\rm(geo)\pm}$ is also shown.
}
\label{fig:velocity}
\end{figure}

Of course as one increases $\zeta^{-1}$ from 0 at a given radius, the 4-velocity along the orthogonal direction to the 2-plane of the rotation goes null and then spacelike as one encounters the direction at which the rotation becomes a null rotation and then a boost. Since for geodesics their own 4-velocity is parallel transported, one has $\mathcal{Z}(\zeta_{\rm(geo)\pm}) =\zeta_{\rm(geo)\pm}$ and the 2-plane of the rotation lies in the local rest space of the geodesics themselves. Thus the orbit angular velocity interval where this 2-plane is not spacelike lies somewhere between the two geodesic angular velocities $\zeta_{\rm(geo)\pm}$ when they are both timelike. In fact since $\mathcal{Z}^2=\mathcal{Z}$, the endpoints are $\mathcal{Z}(\zeta_\pm)$, where $\zeta_\pm$ are the angular velocities of the pair of oppositely rotating null circular orbits at a given radius.
As shown in Fig.~\ref{fig:velocity}, moving towards the black hole, this interval 
$[\mathcal{Z}(\zeta_-),\mathcal{Z}(\zeta_+)] $
expands until it first encounters on its left side the counterrotating geodesic at the radius where it goes null, while even closer to the hole it then encounters on its right side the corotating geodesic at the radius where it goes null. This interval continues expanding until the radius $r_{\rm(gmp)}$ of the GMPO horizon, where the left endpoint becomes infinite, corresponding to the null rotation which occurs for the closed $\phi$ loops at that radius, while the right endpoint finally goes to infinity at the black hole horizon. The left endpoint then re-enters the top right corner of the plot to terminate at the horizon again at infinite velocity. 

To explicitly construct a parallel transported orthonormal frame along one of these circular orbits, say for an angular velocity $\zeta$ outside the boost zone 
at a radius outside the GMPO horizon, 
one can take $u(\mathcal{Z}(\zeta))$ and $e_{\hat\theta}$, which are orthogonal and both parallel transported along the orbit, and the pair of orthonormal unit vectors in the orthogonal 2-plane which result from the parallel transport of the vectors
$e_{\hat r}$ and $e_{\hat\phi}(\mathcal{Z}(\zeta))$ from their initial values through an angle $\Phi$ (in the counterclockwise direction) related to the change in $\phi$ (in the clockwise direction) from its initial value by the constant ratio (\ref{eq:Phiphi}).
Inside the boost zone $u(\mathcal{Z}(\zeta)$ and $e_{\hat\phi}(\mathcal{Z}(\zeta))$ simply swap causality properties and a boost replaces the rotation. On the boost zone boundary where $u(\mathcal{Z}(\zeta)$ is null, a little more effort is required to get such an orthonormal frame. 

\section{Fermi-Walker transport}

Finally one can construct the corresponding Fermi frames along the timelike  circular orbits by considering the relationship between parallel transport and Fermi-Walker transport. If one considers the spacetime Fermi-Walker transport equation (\ref{eq:fwtransport}) applied to a vector $X$ which is orthogonal to $u$, i.e. `spatial', then its Fermi-Walker derivative is also 
spatial so spatially projecting the equation leads to
\beq\label{eq:sfwtransport}
\frac{D_{\rm(fw)} X}{ds}
=P(u)\frac{D_{\rm(fw)} X}{ds}
=P(u)\frac{D X}{ds}\ .
\eeq
Thus for spatial vectors, Fermi-Walker transport is just spatially projected parallel transport at the derivative level. 

Projecting the covariant derivative of the vector field $e_{\hat r}$ (given by Eq.~(\ref{eq:Der2})) into the local rest space of 
$u(\zeta)$ 
yields its Fermi-Walker derivative along $u(\zeta)$ 
\begin{eqnarray}\label{eq:PDer}
  \frac{D_{\rm(fw)} e_{\hat r}}{d\phi}  
  &=& P(U(\zeta))\frac{D e_{\hat r}}{d\phi}
=  \frac{d\Phi}{d\phi} P(u(\zeta))\, e_{\hat\phi}(\mathcal{Z}(\zeta))
\nonumber\\
&=& \frac{d\Phi}{d\phi} \gamma(u(\mathcal{Z}(\zeta)),u(\zeta)) \, e_{\hat\phi}(\zeta)\ .
\end{eqnarray}
Thus the ratio between the Fermi-Walker transport angle in the plane of
$e_{\hat r}$ and $e_{\hat\phi}(\zeta)$ of the local rest space of 
$u(\zeta)$
and the orbital angle is instead
\beq\label{eq:Phiphifw}
  \frac{d\Phi_{\rm(fw)}(\zeta)}{d\phi}
   = \gamma(u(\mathcal{Z}(\zeta)),u(\zeta)) 
          \,\frac{d\Phi(\zeta)}{d\phi}\ ,
\eeq
with an extra relative gamma factor describing the inverse Lorentz contraction which occurs in the projection, which has unit value for geodesics where this projection reduces to the identity.
The relative gamma factor may be expressed in terms of the ZAMO gamma factors \cite{rok} as 
\beq
\gamma(u(\mathcal{Z}(\zeta)),u(\zeta))
= \gamma(\mathcal{Z}(\zeta)) \gamma(\zeta)
  [1-\nu^{\hat\phi}(\mathcal{Z}(\zeta)) \nu^{\hat\phi}(\zeta)]\ .
\eeq

This extra gamma factor for Fermi-Walker transport relative to parallel transport allows $  D\Phi_{\rm(fw)}(\zeta)/d\phi>1$, which can then lead to a retrograde rotation rather than a prograde rotation as in the parallel transport case. Indeed in the flat spacetime limit where $\mathcal{Z}(\zeta)=0$ (no tilt of the parallel transport plane relative to the Minkowski space time coordinate hypersurfaces), $\zeta_{\rm(gmp)}=0$ (the limiting GMPOs follow the time coordinate lines) and $R=\rho=r$ 
(no spatial curvature), only this factor $\gamma(u(0),u(\zeta))=\gamma(\zeta)$ remains and leads to the retrograde Thomas precession in the local rest space of the circular orbit with angular frequency 
$ d\Phi_{\rm(fw)}(\zeta)/d\phi-1 =\gamma(\zeta)-1$
described explicitly in exercise 6.9 of Misner, Thorne and Wheeler.\cite{mtw}

To explicitly construct a Fermi-Walker transported frame along one of these circular orbits, say for an angular velocity $\zeta$ outside the boost zone at a radius outside the GMPO horizon, one can take $e_0=u(\zeta)$ and $e_3=-e_{\hat\theta}$, which are orthogonal and Fermi-Walker transported along the orbit, and the pair of orthonormal unit vectors $e_1,e_2$ in the orthogonal 2-plane which result from the Fermi-Walker transport of the vectors
$e_{\hat r}$ and $e_{\hat\phi}(\zeta)$ respectively from their initial values through an angle $\Phi_{\rm(fw)}$ (in the counterclockwise direction) related to the change in $\phi$ (in the clockwise direction) from its initial value by the constant ratio (\ref{eq:Phiphifw}). 

\begin{figure}
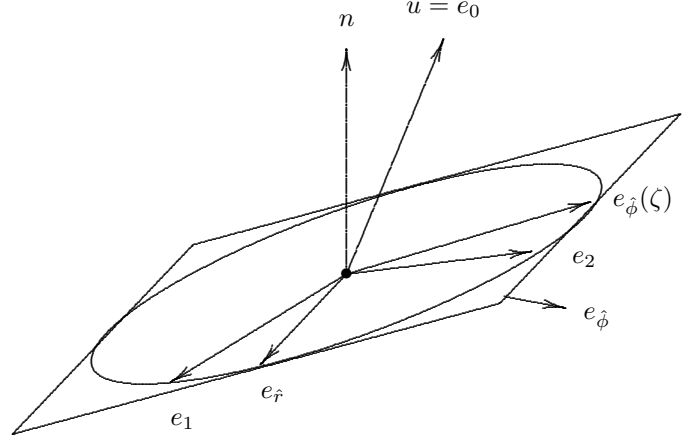

\typeout{figure 8}
\vglue -4cm
$$ \vbox{
\beginpicture
\setcoordinatesystem units <3cm,3cm> point at  0 10  

\put {\mathput{\bullet}}              at  0 0
\put {\mathput{n}}     [b]            at  0 1.1
\put {\mathput{u=e_0}}     [b]            at  0.43 1.14
\put {\mathput{e_2}}  [lt]            at  1.0 0.10
\put {\mathput{e_{\hat\phi}(\zeta)}}  [l] at  1.18 0.32
\put {\mathput{e_{\hat\phi}}}  [l]    at  1.05 -0.2
\put {\mathput{e_{\hat r}}}  [tl]     at  -0.38 -0.50
\put {\mathput{e_1}}  [tl]            at  -0.78 -0.62


\arrow <.3cm>  [.1,.4]    from  0 0 to 0 1 
\arrow <.3cm>  [.1,.4]    from  0 0 to 0.43 1.04
\arrow <.3cm>  [.1,.4]    from  0 0 to 1.08 0.32
\arrow <.3cm>  [.1,.4]    from  0 0 to 0.82 0.10
\arrow <.3cm>  [.1,.4]    from  0.70 -0.10 to 0.97 -0.15
\arrow <.3cm>  [.1,.4]    from  0 0 to -0.38 -0.40
\arrow <.3cm>  [.1,.4]    from  0 0 to -0.78 -0.48 

\plot -0.68 0.13  1.48 0.71 0.68 -0.13 -1.48 -0.71  -0.68 0.13 / 
\setplotarea x from -1.2 to 1.2, y from -2 to 2
\startrotation by .93969 .34202
\ellipticalarc axes ratio 120:28 360 degrees from 1.2 0 center at 0 0
\endpicture}$$
\vglue -4cm
\caption{The Minkowski spacetime Fermi frame with $e_3=-e_{\hat\theta}$ suppressed. 
The local rest space of $u$ is shown relative to the nonrotating (parallel transported) time direction $n$ and its local rest space containing the vectors $e_{\hat r}$ and $e_{\hat\phi}$. The Fermi frame vectors $e_1$ and $e_2$ undergo a counterclockwise rotation relative to the boosted axes 
$e_{\hat r}$ and $e_{\hat\phi}(\zeta)$.
}
\label{fig:fermiframe}
\end{figure}

With the abbreviations $\gamma=\gamma(\zeta)$ and $\nu^{\hat\phi}=\nu^{\hat\phi}(\zeta)$, one   has
\beq
  e_0 = u(\zeta) = \gamma [n +\nu^{\hat\phi}e_{\hat\phi}]\ ,\qquad
  e_{\hat\phi}(\zeta) = \gamma [e_{\hat\phi} +\nu^{\hat\phi}n]\ ,
\eeq
or
\beq
  e_0 \pm e_{\hat\phi}(\zeta) = e^{\pm\alpha} [n\pm e_{\hat\phi}]\ , 
\eeq
where $\nu^{\hat\phi} = \tanh \alpha$. 
Next, letting $e_\pm=e_1\pm i e_2$, one has to apply the Fermi rotation to the $e_{\hat r}$-$e_{\hat\phi}(\zeta)$ plane
\beq
 e_+= e_1 + i e_2 = e^{i\Phi_{\rm(fw)}} [ e_{\hat r} + i e_{\hat\phi}(\zeta)]\ ,
\eeq
where one can take $\Phi_{\rm(fw)} = (D\Phi_{\rm(fw)}/d\phi) \, \phi$ along the orbit $\phi=\zeta t +\phi_0$. Letting $e_3=-e_{\hat\theta}$,
then ${e_0,e_1,e_2,e_3}$ is a spatially righthanded Fermi frame, illustrated in Fig.~\ref{fig:fermiframe} by suppressing the $\theta$ direction.

In the flat spacetime case, one has $\Phi_{\rm(fw)}=\gamma \phi$ and
$\nu^{\hat\phi}=r\zeta$. This frame was apparently first given by
Pirani\cite{pirani2} (1957), as well as for the case of circular geodesics in the Schwarzschild spacetime, and later was used by Irvine\cite{irvine} (1964) and Corum\cite{corum} (1980) to study Maxwell's equations in a uniformly rotating coordinate system in Minkowski spacetime.

\begin{figure}
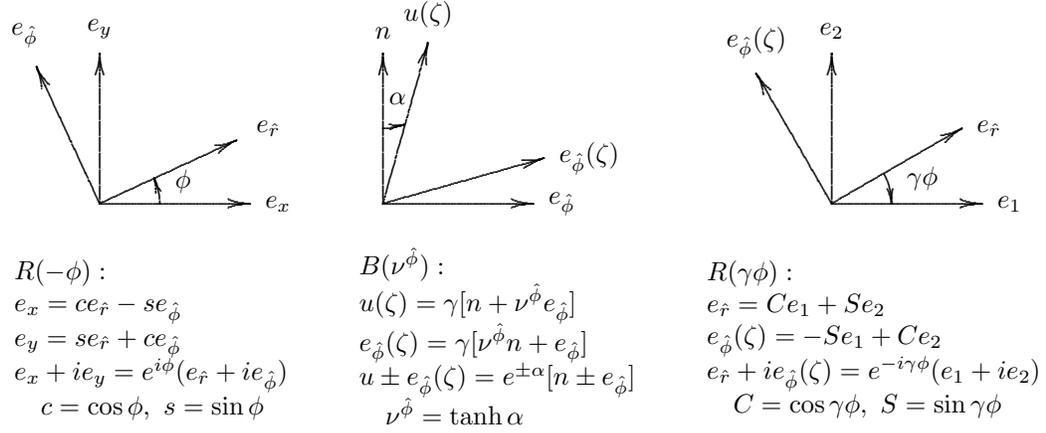

\typeout{*** Figure 9}
\begin{tabular}{ccc}
\vbox{
\beginpicture
\setcoordinatesystem units <2cm,2cm> point at  0 0  
\put {\mathput{e_y}} [b]                at  0 1.1
\put {\mathput{e_x}} [l]                at  1.1 0
\put {\mathput{e_{\hat r}}} [l]         at  1.04 0.5
\put {\mathput{e_{\hat\phi}}} [b]       at -0.5 1.04 
\put {\mathput{\phi}} [l]                at  0.5 0.15
\arrow <.3cm>  [.1,.4]    from  0 0 to 0 1 
\arrow <.3cm>  [.1,.4]    from  0 0 to 1 0
\arrow <.3cm>  [.1,.4]    from  0 0 to 0.91 0.42
\arrow <.3cm>  [.1,.4]    from  0 0 to -0.42 .91
\arrow <.2cm>  [.1,.4]    from  0.374 0.129 to 0.362 0.169
\circulararc 25 degrees from 0.4 0 center at 0 0
\endpicture}
&
\vbox{
\beginpicture
\setcoordinatesystem units <2cm,2cm> point at  0 0  
\put {\mathput{n}} [b]                at  0 1.1
\put {\mathput{e_{\hat\phi}}} [l]     at  1.1 0
\put {\mathput{u(\zeta)}} [b]                at  0.3 1.17
\put {\mathput{e_{\hat\phi}(\zeta)}} [l] at  1.17 0.3
\arrow <.3cm>  [.1,.4]    from  0 0 to 0 1 
\arrow <.3cm>  [.1,.4]    from  0 0 to 1 0
\arrow <.3cm>  [.1,.4]    from  0 0 to 1.07 0.3
\arrow <.3cm>  [.1,.4]    from  0 0 to 0.3 1.07
\arrow <.2cm>  [.1,.4]    from  0.11 0.52  to 0.14 0.525 
\circulararc 16.5 degrees from 0 0.5 center at 0 1
\put {\mathput{\alpha}} [l]                at  0.04 0.7
\endpicture}
&
\vbox{
\beginpicture
\setcoordinatesystem units <2cm,2cm> point at  0 0  
\put {\mathput{e_2}} [b]                at  0 1.1
\put {\mathput{e_1}} [l]                at  1.1 0
\put {\mathput{e_{\hat r}}} [l]          at  .966 0.5
\put {\mathput{e_{\hat\phi}(\zeta)}} [b] at -0.5 .966 
\arrow <.3cm>  [.1,.4]    from  0 0 to 0 1 
\arrow <.3cm>  [.1,.4]    from  0 0 to 1 0
\arrow <.3cm>  [.1,.4]    from  0 0 to 0.866 0.5
\arrow <.3cm>  [.1,.4]    from  0 0 to -0.5 .866
\circulararc 30 degrees from 0.4 0 center at 0 0
\arrow <.2cm>  [.1,.4]    from  0.395 .04 to 0.4 0
\put {\mathput{\gamma\phi}} [l]                at  0.5 0.17
\endpicture}
\\
\null& & \\
\begin{tabular}{c}
$\begin{array}{l}
R(-\phi):\\
e_x = c e_{\hat r} - s e_{\hat\phi}\\
e_y = s e_{\hat r} + c e_{\hat\phi}\\
e_x+ i e_y = e^{i\phi} (e_{\hat r}+ie_{\hat\phi})\\
\quad c=\cos\phi,\ s=\sin\phi 
\end{array}$
\end{tabular}
&%
\begin{tabular}{c}
$\begin{array}{l}
B(\nu^{\hat\phi}):\\
u(\zeta) = \gamma[ n + \nu^{\hat\phi} e_{\hat\phi} ]\\
e_{\hat\phi}(\zeta) = \gamma[\nu^{\hat\phi} n +  e_{\hat\phi} ]\\
u\pm e_{\hat\phi}(\zeta) 
       = e^{\pm\alpha}[ n \pm  e_{\hat\phi} ]\\
\quad \nu^{\hat\phi}=\tanh\alpha 
\end{array}$
\end{tabular}
&%
\begin{tabular}{c}
$\begin{array}{l}
R(\gamma\phi):\\
e_{\hat r} = C e_1 + S e_2\\
e_{\hat\phi}(\zeta) = -S e_1 +C e_2\\
e_{\hat r} + i e_{\hat\phi}(\zeta) = e^{-i\gamma\phi} (e_1+ie_2)\\
\quad C=\cos\gamma\phi,\ S=\sin\gamma\phi 
\end{array}$
\end{tabular}
\end{tabular}
\caption{The successive transformations from the parallel transported frame to the Fermi frame in Minkowski spacetime.}
\label{fig:RBR}
\end{figure}

If $n,e_x,e_y,e_z=-e_{\hat\theta}$ is the nonrotating orthonormal inertial coordinate frame in Minkowski spacetime (at $\theta=\pi/2$), and therefore parallel transported along any curve, then it can be expressed in terms of the Fermi frame along a circular orbit with angular frequency $\zeta$
by a rotation by angle $\gamma\phi$ in the $e_1$-$e_2$ plane, followed by a boost with velocity $-\nu^{\hat\phi}$ in the $\phi$ direction, followed by a rotation by the angle $-\phi$ in the  $e_{\hat r}$-$e_{\hat\phi}$ plane
\beq
     \left[ \begin{array}{c} n\\ e_x\\ e_y\\ e_z\end{array}\right]
    = R(-\phi) B(-\nu^{\hat\phi}) R(\gamma\phi) \,
     \left[ \begin{array}{c} u\\ e_1\\ e_2\\ e_3\end{array}\right]\ .
\eeq
This is illustrated in Fig.~\ref{fig:RBR}.

The result of these three transformations, is
\begin{eqnarray}
e_2 &=& e_3\ ,\quad
n = \gamma [u -\nu^{\hat\phi} \hbox{Im}(e^{-i\gamma\phi} e_+)]\ ,\nonumber\\
e_x+i e_y &=& e^{-i(\gamma-1)\phi} e_+ 
      + i (\gamma-1) e^{i\gamma\phi} \hbox{Im}(e^{-i\phi} e_+)
      - i e^{i\gamma\phi} \gamma \nu^{\hat\phi} u\ .
\end{eqnarray}
If
\beq
   S = S^n n + P(n)S\ ,\ S\cdot u = 0\ ,\ S^n = -S\cdot n
\eeq
is the spin vector of a test gyroscope carried along the circular orbit, belonging to the local rest space of the orbit, then its `laboratory' components, letting $S_\pm = S\cdot e_\pm$, $S_x = S\cdot e_x$, etc., are
\begin{eqnarray}
 S_z&=& S_3\ ,\quad S^n = - \gamma \nu^{\hat\phi} \hbox{Im}(e^{-i\gamma\phi} S_+)\ ,\nonumber\\
S_x+ i S_y &=& e^{-i(\gamma-1)\phi} S_+ 
      + i (\gamma-1) e^{i\gamma\phi} \hbox{Im}(e^{-i\phi} S_+)\ .
\end{eqnarray}
Note that the Fermi frame components are constants since the spin vector is Fermi-Walker transported. The case $S_2=0$, $S_+=S_1=S_3$ reproduces Eqn.~(6.28) of Misner, Thorne Wheeler.\cite{mtw}

The first term on the right hand side of the last equation, together with $S_z$, is the boosted spin vector $B(n,u)S$ actively boosted back from the local rest space of the circular orbit to the local rest space of the nonrotating laboratory frame, while the second term is the distortion in the spin vector due to its relative motion.
The active boost identifies $e_{\hat\phi}(\zeta)$ and $e_{\hat\phi}$, removing the passive boost sandwiched between the two passive rotations which simply re-expresses the frames in terms of each other, leading to the pure rotation $R((\gamma-1)\phi)$ of the laboratory boosted spin vector compared to the laboratory axes with the Thomas precession angular velocity 
\beq
(\gamma-1)\zeta = \frac{\gamma^2\nu^2}{\gamma+1} \zeta\ ,
\eeq 
a secular precession which grows with time. The distortion term is a mere periodic oscillation which averages out in time. 

\begin{figure}
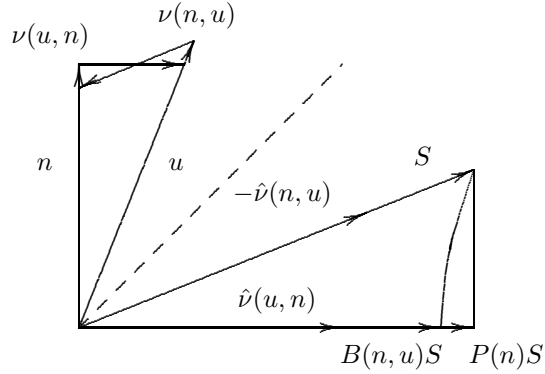

\typeout{*** Figure 10}
$$ \vbox{
\beginpicture
  \setcoordinatesystem units <.7cm,.7cm> point at 0 0  

    \putrule from 0 0   to 0 5.0 
    \putrule from 0 0   to 7.5 0  
    \putrule from 0 5.0 to 2.0 5.0 
    \putrule from 7.5 0 to 7.5 3.0 

  \setlinear
    \plot 0 0 2.18  5.45  /  
    \plot 0 0 7.5 3.0  /     
    \plot 0 4.55 2.18 5.45 / 

\setdashes
    \plot 0 0 5 5 /  


\setsolid
\setquadratic
  \plot 6.875 0  6.95 0.9  7.05 1.5  7.15 1.9  7.5 3.0 /


\setlinear 
    \arrow <.3cm> [.1,.4]    from  0 4.7 to 0 5.0           
    \arrow <.3cm> [.1,.4]    from  2.06 5.15 to 2.18 5.45   
    \arrow <.3cm> [.1,.4]    from  1.7 5.0 to 2.0 5.0       
    \arrow <.3cm> [.1,.4]    from  0.3 4.7 to 0 4.55        
 
    \arrow <.3cm> [.1,.4]    from  4.7 0 to 5.0 0           
    \arrow <.3cm> [.1,.4]    from  5.15 2.06 to  5.45 2.18  
    \arrow <.3cm> [.1,.4]    from  7.2 2.88 to  7.5 3.0     

    \arrow <.3cm> [.1,.4]    from  6.575 0 to 6.875 0       
    \arrow <.3cm> [.1,.4]    from  7.2 0 to 7.5 0           

  \put {\mathput{n}}                          [rB]   at  -.5 3.0
  \put {\mathput{u}}                          [lB]   at  1.7 3.0
  \put {\mathput{\nu(u,n)}}                   [rB]   at  0.2 5.4
  \put {\mathput{\nu(n,u)}}                   [lB]   at  1.5 5.8

  \put {\mathput{S}}                          [rB]   at  6.7 3.1
  \put {\mathput{\hat\nu(u,n)}}               [rB]   at  4.5 0.4
  \put {\mathput{-\hat\nu(n,u)}}              [rB]   at  4.8 2.4

  \put {\mathput{P(n)S}}   [lt]   at  7.4   -.3
  \put {\mathput{B(n,u)S}}     [rt]   at  6.9 -0.3


\endpicture}$$

\caption{The relationship between the projection and boost between local rest spaces along the direction of relative motion. The relative velocity $\nu(u,n)$ of $n$ with respect to $u$ and $\nu(u,n)$ of $u$ with respect to $n$ are related to each other by the relative boost and a sign change.
The hatted vectors are the corresponding unit vectors.  
If $S$ is along the direction of relative motion in the local rest space of $u$, then since $B(n,u)S = \gamma^{-1} P(n)S$, it follows that
$P(n)S = \gamma B(n,u)S = B(n,u)S + (\gamma-1) B(n,u)S$. 
For an arbitrary vector $S$ in the local rest space of $u$, this figure applies to its component 
$[S\cdot \hat\nu(n,u)] \hat\nu(n,u) $ along $-\hat\nu(n,u)$, which satisfies $B(n,u)[-\hat\nu(n,u)]=\gamma \hat\nu(u,n)$, while the components of $S$ orthogonal to the plane of $n$ and $u$ are unchanged by the boost or projection, leading to the general relationship (\ref{eq:BP}) of the text.
}
\label{fig:BP}
\end{figure}

This decomposition of the measured spatial vector is valid in general
\beq\label{eq:BP}
   P(n)S = B(n,u)S + [\gamma(n,u)-1] [-\hat\nu(n,u)\cdot S] \hat\nu(u,n)
\eeq
and is illustrated in Fig.~\ref{fig:BP} and discussed in Refs.~\citen{mfg,rok}. Long term secular effects can be described by the boosted spin  vector, which precesses with a constant angular velocity, while the second term is a periodic oscillation analogous to the stellar aberration effect for  null  directions. In fact in a gyroscope experiment, with the fixed stars as the reference, it is the spin vector boosted into the local rest space of the static observers which gives the secular spin precession formula (subtracting out the stellar aberration effect), since it is the static observers which are locked onto the distantly nonrotating observers, i.e., the fixed stars. This is calculated exactly for black hole spacetimes in Ref.~\citen{idcf2}. It is exactly this factor of $\gamma(n,u)-1$ which occurs in the projection compared to the boost which leads to the relative (difference) precession frequency itself from Eq.~(\ref{eq:PDer}).

The relationship between the parallel transported and Fermi-Walker transported axes remains true for circular orbits in black hole or even more general stationary axisymmetric spacetimes if for the parallel transported frame one takes $u(\mathcal{Z}(\zeta))$ in place of $n$ and  
$e_{\hat\phi}(\mathcal{Z}(\zeta))$ in place of $e_{\hat\phi}$ 
in the above discussion, with $\gamma$ replaced by the relative gamma factor $\gamma(u(\zeta),u(\mathcal{Z}(\zeta)))$. The secular precession of Fermi-Walker transport compared to parallel transport is governed by the same frequency relationship
\beq
\frac{d\Phi_{\rm(fw)}(\zeta)}{dt} - \frac{d\Phi(\zeta)}{dt}
= [\gamma(u(\zeta),u(\mathcal{Z}(\zeta)))-1]\zeta \ ,
\eeq
for which the Thomas precession is a special case. 
This relationship should remain valid for helical motion in stationary cylindrically symmetric spacetimes as well.

By considering the natural Frenet-Seret frame along a circular orbit (see Ref.~\citen{circfs} and the appendix of \citen{bjm2002}), with curvature $\kappa$ (magnitude of the acceleration in the timelike case) and first torsion $\tau_1$, while the second torsion $\tau_2$ vanishes in the equatorial plane, one easily sees the direct relationship between $u(\mathcal{Z}(\zeta))$ and  $u(\zeta)$ can be expressed through the their relative velocity
\beq
 \nu^{\hat\phi}(u(\mathcal{Z}(\zeta)),u(\zeta))
=\frac{\kappa}{\tau_1}
=\frac{1}{\nu_{gmp}}\frac{(\nu-\nu_+)(\nu-\nu_-)}{(\nu-\nu_{crit+})(\nu-\nu_{crit-})}\ .
\eeq
Here the abbreviations $\nu_\pm$ denote the pair of geodesic relative velocities (zeros of $\kappa$), while $\nu_{crit\pm}$ denote the corresponding critical velocities\cite{idcf2} for $\kappa$ and 
$\nu_{gmp}$ the GMPO relative velocity, all with respect to the ZAMOs. 
This formula follows from formulas (A.5) of Ref.~\citen{bjm2002} for $\tau_1$ and formulas
(4.7) and (4.8) of Ref.~\citen{idcf2}. It can be directly expressed in terms of the relative velocities with respect to the circular orbit itself as follows
\beq
\nu(u({{\mathcal Z}(\zeta)}), u(\zeta)) =2\left(
\frac{1}{\nu(u(\zeta_{\rm geo +}), u(\zeta))}+
\frac{1}{\nu(u(\zeta_{\rm geo -}), u(\zeta))}
\right)^{-1} \ .
\eeq
Since the reciprocal map is the bar map giving the relative velocity of the orthogonal direction in the circular orbit cylinder, i.e. the angular direction for a timelike orbit, this states that the relative velocity of the direction along this cylinder giving the orientation of the Lorentz transformation plane is the average of the relative velocities of the local rest space angular directions of the pair of geodesics (when they are timelike).

The extremely accelerated observers have relative velocity $\nu_{crit-}$ and not only see the timelike circular geodesics with the same relative speed but opposite directions,
but also see the entire boost zone symmetrically.\cite{bjm2002} One can show that the map
$\mathcal{Z}$ when expressed in terms of relative velocities with respect to these observers (tilde notation) takes the simple form
\beq
 \tilde\nu \to -\tilde\nu_+\tilde\nu_-/\tilde\nu 
\quad\hbox{(where}\ \tilde\nu_-=-\tilde\nu_+ \,)
\eeq
characteristic of nonrotating black hole spacetimes, which implies that the boost zone endpoint velocities then become $\tilde\nu(\mathcal{Z}(\zeta_\pm))=\mp \tilde\nu_+\tilde\nu_-$.
Thus parallel transport along circular orbits links together in various ways not only the GMPOs, Carter observers and static observers but also the extremely accelerated observers.


\section{Circular Holonomy and Clock Effects?}

What does all of this have to do with circular holonomy and clock effects?
Rotation in black hole spacetimes introduces an asymmetry between the corotating and counterrotating circular geodesics. The various clock effects measure the difference between the two periods as seen by a given circularly rotating observer for one circuit compared to that observer (either the proper time periods measured by the geodesics themselves, or the observer proper-time periods). Choosing the observers to be the GMPOs leads to the observer-independent clock effect measuring the difference in the geodesic proper periods between every second geodesic crossing point. While all this clock time comparison is going on, it is reasonable to compare gyro spins as well to see how the rotation effects these differently on the pair of geodesics.

Holonomy is the study of how curvature affects vectors during parallel transport around closed loops from a fixed reference point; since parallel transport preserves length, only the direction can change by an element of the Lorentz group with respect to a fixed orthonormal frame in the tangent space at the reference point. The holonomy group at a given point is the subgroup of the Lorentz group which contains all the Lorentz transformations which result from all possible loops starting and stopping at that point. If one restricts the loops to piecewise smooth stationary circular orbits at a fixed radius in the equatorial plane of a black hole, one explores a subset of the holonomy group. The simplest such closed loops are the closed $\phi$ loops characterized completely by the integer number $q$ of corotating $q>0$ or counterrotating $q<0$ circuits. 

One can ask when this discrete set of holonomy Lorentz transformations contains the identity. Since parallel transport induces a one-parameter family of such transformations along the orbit relative to the orthonormal spherical frame, the radius must be in the rotation zone in order for these transformations to return to the identity. The rotation zone corresponds to where the GMPOs are timelike,  a zone which terminates at their horizon approaching the black hole where $\gamma(\zeta_{\rm(mgp)})\to\infty$. Then
\beq
      \frac{d\Phi}{d\phi} (2\pi q ) = 2\pi p \rightarrow
      \frac{d\Phi}{d\phi}  = p/q 
\eeq
shows that in this zone where $0<d\Phi/d\phi<1$, for each proper fractional rational number value of this relative frequency function (which occurs at a dense set of radii in this zone), parallel transport will return every vector to its original state after a certain number of loops, leading to what Rothman, Ellis and Murugan\cite{rothman} have called a band of holonomy invariance extending from the GMPO horizon out to infinity. For black hole spacetimes only, this relative frequency function turns out to be the ratio of the proper (self) period of the GMPOs for one $\phi$ coordinate loop to the average coordinate time period of the pair of oppositely rotating circular geodesics for the same loop, i.e., the time as seen by the distantly nonrotating observers far from the hole.

One can consider a stationary circular orbit which does not close but these are helices in spacetime so one needs two such orbits  joined together at an initial point to obtain closed circuits. If one takes one to be one of the various geometrically preferred observers which generalize different aspects of the static spacetime nonrotating observers and the other a timelike geodesic, one can compare how a vector transported around the hole on a circular geodesic compares to the one carried by the observer when they meet. Or one can take a pair of oppositely rotating timelike circular geodesics which start at a common initial point. In the latter case for black holes only, the relative frequency function for each geodesic is also a simple ratio of the proper period (for a circuit defined by the geodesic crossing points themselves) to the average coordinate period for one $\phi$ coordinate loop.

However, parallel transported vectors along timelike curves are not directly connected with an interesting physical property, while Fermi-Walker transported vectors describe how test gyros behave along these world lines, so it makes sense to extend the notion of holonomy to Fermi-Walker transport in order to compare how gyro spin vectors differ along different circular orbits from the same initial point when they meet again. This requires allowing a relative boost to identify corresponding spin vectors at meeting points since the spin vectors lie in distinct local rest spaces in general. This leads to `spin holonomy'. For clock effect oppositely-rotating timelike circular geodesic pairs, this discussion also involves the clock effect periods.\cite{bjm2002,mash}

Circular orbits have grabbed the imaginations of so many of us over the past century. The present discussion has shown that their geometric richness has still not yet been depleted and has led to further insight about Fermi-Walker transport itself in this context.


\section*{Acknowledgements}

Professor Luigi Stazi is thanked for his help in obtaining Levi-Civita's articles from the library of the Mathematics Department of the University of Rome ``La Sapienza" where Levi-Civita  taught for many years.

\end{document}